\input harvmac
\let\includefigures=\iftrue
\let\useblackboard==\iftrue
\newfam\black

\includefigures
\message{If you do not have epsf.tex (to include figures),}
\message{change the option at the top of the tex file.}
\input epsf
\def\figin{\epsfcheck\figin}\def\figins{\epsfcheck\figins}
\def\epsfcheck{\ifx\epsfbox\UnDeFiNeD
\message{(NO epsf.tex, FIGURES WILL BE IGNORED)}
\gdef\figin##1{\vskip2in}\gdef\figins##1{\hskip.5in}
\else\message{(FIGURES WILL BE INCLUDED)}%
\gdef\figin##1{##1}\gdef\figins##1{##1}\fi}
\def\DefWarn#1{}
\def\figinsert{\goodbreak\midinsert}
\def\ifig#1#2#3{\DefWarn#1\xdef#1{fig.~\the\figno}
\writedef{#1\leftbracket fig.\noexpand~\the\figno}%
\figinsert\figin{\centerline{#3}}\medskip\centerline{\vbox{
\baselineskip12pt\advance\hsize by -1truein
\noindent\footnotefont{\bf Fig.~\the\figno:} #2}}
\endinsert\global\advance\figno by1}
\else
\def\ifig#1#2#3{\xdef#1{fig.~\the\figno}
\writedef{#1\leftbracket fig.\noexpand~\the\figno}%
\global\advance\figno by1} \fi

\def\id{{1 \kern-.28em {\rm l}}}

\def\Z{{\bf Z}}

\def\K3{{\bf K3}}
\def\journal#1&#2(#3){\unskip, \sl #1\ \bf #2 \rm(19#3) }
\def\andjournal#1&#2(#3){\sl #1~\bf #2 \rm (19#3) }

\def\bar{\overline}
\def\hat{\widehat}
\def\ie{{\it i.e.}}
\def\eg{{\it e.g.}}

\def\etc{{\it etc}}

\def\tilde{\widetilde}

\def\frac#1#2{{#1\over#2}}

\def\inbar{\,\vrule height1.5ex width.4pt depth0pt}
\def\IC{\relax\hbox{$\inbar\kern-.3em{\rm C}$}}
\def\IR{\relax{\rm I\kern-.18em R}}
\def\IP{\relax{\rm I\kern-.18em P}}
\def\Z{{\bf Z}}

%
%

%
\catcode`\@=11
\def\slash#1{\mathord{\mathpalette\c@ncel{#1}}}
\overfullrule=0pt

\def\underrel#1\over#2{\mathrel{\mathop{\kern\z@#1}\limits_{#2}}}

\catcode`\@=12


%

\def\tr{{\rm tr}}

\def \sinh{{\rm sinh}}
\def \cosh{{\rm cosh}}


\def\ie{{\it i.e.}}
\def\eg{{\it e.g.}}


\lref\Zamolodchikov{
A.~B.~Zamolodchikov,
``Expectation value of composite field T anti-T in two-dimensional quantum field theory,''
[arXiv:hep-th/0401146 [hep-th]].
}

\lref\Smirnov{
F.~A.~Smirnov and A.~B.~Zamolodchikov,
``On space of integrable quantum field theories,''
Nucl. Phys. B {\bf 915}, 363-383 (2017)
doi:10.1016/j.nuclphysb.2016.12.014
[arXiv:1608.05499 [hep-th]].
}

\lref\Cavaglia{
A.~Cavagli\`a, S.~Negro, I.~M.~Sz\'ecs\'enyi and R.~Tateo,
``$T \bar{T}$-deformed 2D Quantum Field Theories,''
JHEP {\bf 10}, 112 (2016)
doi:10.1007/JHEP10(2016)112
[arXiv:1608.05534 [hep-th]].
}

\lref\Giveon{
A.~Giveon, N.~Itzhaki and D.~Kutasov,
``$T{\bar T} $ and LST,''
JHEP {\bf 07}, 122 (2017)
doi:10.1007/JHEP07(2017)122
[arXiv:1701.05576 [hep-th]].
}

\lref\Hassan{
S.~F.~Hassan and A.~Sen,
``Marginal deformations of WZNW and coset models from O(d,d) transformation,''
Nucl. Phys. B {\bf 405}, 143-165 (1993)
doi:10.1016/0550-3213(93)90429-S
[arXiv:hep-th/9210121 [hep-th]].
}

\lref\Boyda{
E.~K.~Boyda, S.~Ganguli, P.~Horava and U.~Varadarajan,
``Holographic protection of chronology in universes of the Godel type,''
Phys. Rev. D {\bf 67}, 106003 (2003)
doi:10.1103/PhysRevD.67.106003
[arXiv:hep-th/0212087 [hep-th]].
}

\lref\Gauntlett{
J.~P.~Gauntlett, J.~B.~Gutowski, C.~M.~Hull, S.~Pakis and H.~S.~Reall,
``All supersymmetric solutions of minimal supergravity in five- dimensions,''
Class. Quant. Grav. {\bf 20}, 4587-4634 (2003)
doi:10.1088/0264-9381/20/21/005
[arXiv:hep-th/0209114 [hep-th]].
}

\lref\Brace{
D.~Brace, C.~A.~R.~Herdeiro and S.~Hirano,
``Classical and quantum strings in compactified pp waves and Godel type universes,''
Phys. Rev. D {\bf 69}, 066010 (2004)
doi:10.1103/PhysRevD.69.066010
[arXiv:hep-th/0307265 [hep-th]].
}

\lref\Harmark{
T.~Harmark and T.~Takayanagi,
``Supersymmetric Godel universes in string theory,''
Nucl. Phys. B {\bf 662}, 3-39 (2003)
doi:10.1016/S0550-3213(03)00349-3
[arXiv:hep-th/0301206 [hep-th]].
}

\lref\Berenstein{
D.~E.~Berenstein, J.~M.~Maldacena and H.~S.~Nastase,
``Strings in flat space and pp waves from N=4 superYang-Mills,''
JHEP {\bf 04}, 013 (2002)
doi:10.1088/1126-6708/2002/04/013
[arXiv:hep-th/0202021 [hep-th]].
}

\lref\Araujo{
T.~Araujo, E.~\'O.~Colg\'ain, Y.~Sakatani, M.~M.~Sheikh-Jabbari and H.~Yavartanoo,
``Holographic integration of $T \bar{T}$ \& $J \bar{T}$ via $O(d,d)$,''
JHEP {\bf 03}, 168 (2019)
doi:10.1007/JHEP03(2019)168
[arXiv:1811.03050 [hep-th]].
}

\lref\ApoloTT{
L.~Apolo, S.~Detournay and W.~Song,
``TsT, $T\bar{T}$ and black strings,''
JHEP {\bf 06}, 109 (2020)
doi:10.1007/JHEP06(2020)109
[arXiv:1911.12359 [hep-th]].
}

\lref\Godel{
K.~Godel,
``An Example of a new type of cosmological solutions of Einstein's field equations of graviation,''
Rev. Mod. Phys. {\bf 21}, 447-450 (1949)
doi:10.1103/RevModPhys.21.447
}

\lref\Lunin{
O.~Lunin and J.~M.~Maldacena,
``Deforming field theories with U(1) x U(1) global symmetry and their gravity duals,''
JHEP {\bf 05}, 033 (2005)
doi:10.1088/1126-6708/2005/05/033
[arXiv:hep-th/0502086 [hep-th]].
}

\lref\Asrat{
M.~Asrat,
``$T{\bar T}$ and Holography,''
doi:10.6082/uchicago.3365
[arXiv:2112.02596 [hep-th]].
}

\lref\McGough{
L.~McGough, M.~Mezei and H.~Verlinde,
``Moving the CFT into the bulk with $ T\overline{T} $,''
JHEP {\bf 04}, 010 (2018)
doi:10.1007/JHEP04(2018)010
[arXiv:1611.03470 [hep-th]].
}

\lref\Berkovits{
N.~Berkovits, C.~Vafa and E.~Witten,
``Conformal field theory of AdS background with Ramond-Ramond flux,''
JHEP {\bf 03}, 018 (1999)
doi:10.1088/1126-6708/1999/03/018
[arXiv:hep-th/9902098 [hep-th]].
}

\lref\Gotz{
G.~Gotz, T.~Quella and V.~Schomerus,
``The WZNW model on PSU(1,1|2),''
JHEP {\bf 03}, 003 (2007)
doi:10.1088/1126-6708/2007/03/003
[arXiv:hep-th/0610070 [hep-th]].
}

\lref\Buser{
M.~Buser, E.~Kajari and W.~P.~Schleich,
``Visualization of the G\"odel universe,''
New J. Phys. {\bf 15}, 013063 (2013)
doi:10.1088/1367-2630/15/1/013063
[arXiv:1303.4651 [gr-qc]].
}

\lref\Buscher{
T.~H.~Buscher,
``A Symmetry of the String Background Field Equations,''
Phys. Lett. B {\bf 194}, 59-62 (1987)
doi:10.1016/0370-2693(87)90769-6
}

\lref\BuscherB{
T.~H.~Buscher,
``Path Integral Derivation of Quantum Duality in Nonlinear Sigma Models,''
Phys. Lett. B {\bf 201}, 466-472 (1988)
doi:10.1016/0370-2693(88)90602-8
}

\lref\Rocek{
M.~Rocek and E.~P.~Verlinde,
``Duality, quotients, and currents,''
Nucl. Phys. B {\bf 373}, 630-646 (1992)
doi:10.1016/0550-3213(92)90269-H
[arXiv:hep-th/9110053 [hep-th]].
}

\lref\Alvarez{
E.~Alvarez, L.~Alvarez-Gaume, J.~L.~F.~Barbon and Y.~Lozano,
``Some global aspects of duality in string theory,''
Nucl. Phys. B {\bf 415}, 71-100 (1994)
doi:10.1016/0550-3213(94)90067-1
[arXiv:hep-th/9309039 [hep-th]].
}

\lref\Hawking{
S.~W.~Hawking and G.~F.~R.~Ellis,
``The Large Scale Structure of Space-Time,''
Cambridge University Press, 2023,
ISBN 978-1-00-925316-1, 978-1-00-925315-4, 978-0-521-20016-5, 978-0-521-09906-6, 978-0-511-82630-6, 978-0-521-09906-6
doi:10.1017/9781009253161
}

\lref\Osten{
D.~Osten and S.~J.~van Tongeren,
``Abelian Yang-Baxter deformations and TsT transformations,''
Nucl. Phys. B {\bf 915}, 184-205 (2017)
doi:10.1016/j.nuclphysb.2016.12.007
[arXiv:1608.08504 [hep-th]].
}

\lref\BerkovitsB{
N.~Berkovits and J.~Maldacena,
``Fermionic T-Duality, Dual Superconformal Symmetry, and the Amplitude/Wilson Loop Connection,''
JHEP {\bf 09}, 062 (2008)
doi:10.1088/1126-6708/2008/09/062
[arXiv:0807.3196 [hep-th]].
}

\lref\Bekenstein{
J.~D.~Bekenstein,
``Black holes and entropy,''
Phys. Rev. D {\bf 7}, 2333-2346 (1973)
doi:10.1103/PhysRevD.7.2333
}

\lref\Wald{
R.~M.~Wald,
``Black hole entropy is the Noether charge,''
Phys. Rev. D {\bf 48}, no.8, R3427-R3431 (1993)
doi:10.1103/PhysRevD.48.R3427
[arXiv:gr-qc/9307038 [gr-qc]].
}

\lref\Sfondrini{
A.~Sfondrini and S.~J.~van Tongeren,
``$T\bar{T}$ deformations as $TsT$ transformations,''
Phys. Rev. D {\bf 101}, no.6, 066022 (2020)
doi:10.1103/PhysRevD.101.066022
[arXiv:1908.09299 [hep-th]].
}

\lref\Penrose{
R.~Penrose,
``Any Space-Time has a Plane Wave as a Limit,''
Springer Netherlands, Dordrecht, p. 271--27, ISBN 978-94-010-1508-0 (1976)
}

\lref\Gueven{
R.~Gueven,
``Plane wave limits and T duality,''
Phys. Lett. B {\bf 482}, 255-263 (2000)
doi:10.1016/S0370-2693(00)00517-7
[arXiv:hep-th/0005061 [hep-th]].
}

\lref\Freedman{
D.~Z.~Freedman, S.~S.~Gubser, K.~Pilch and N.~P.~Warner,
``Renormalization group flows from holography supersymmetry and a c theorem,''
Adv. Theor. Math. Phys. {\bf 3}, 363-417 (1999)
doi:10.4310/ATMP.1999.v3.n2.a7
[arXiv:hep-th/9904017 [hep-th]].
}

\lref\Girardello{
L.~Girardello, M.~Petrini, M.~Porrati and A.~Zaffaroni,
``Novel local CFT and exact results on perturbations of N=4 superYang Mills from AdS dynamics,''
JHEP {\bf 12}, 022 (1998)
doi:10.1088/1126-6708/1998/12/022
[arXiv:hep-th/9810126 [hep-th]].
}

\lref\Gubser{
S.~S.~Gubser,
``Curvature singularities: The Good, the bad, and the naked,''
Adv. Theor. Math. Phys. {\bf 4}, 679-745 (2000)
doi:10.4310/ATMP.2000.v4.n3.a6
[arXiv:hep-th/0002160 [hep-th]].
}

\lref\Dubovsky{
S.~Dubovsky, T.~Gregoire, A.~Nicolis and R.~Rattazzi,
``Null energy condition and superluminal propagation,''
JHEP {\bf 03}, 025 (2006)
doi:10.1088/1126-6708/2006/03/025
[arXiv:hep-th/0512260 [hep-th]].
}

\lref\Casimir{
H.~B.~G.~Casimir,
``On the Attraction Between Two Perfectly Conducting Plates,''
Indag. Math. {\bf 10}, 261-263 (1948)
}

\lref\Aharony{
O.~Aharony, S.~Datta, A.~Giveon, Y.~Jiang and D.~Kutasov,
``Modular invariance and uniqueness of $T\bar{T}$ deformed CFT,''
JHEP {\bf 01}, 086 (2019)
doi:10.1007/JHEP01(2019)086
[arXiv:1808.02492 [hep-th]].
}

\lref\AsratT{
M.~Asrat,
``KdV charges and the generalized torus partition sum in $T \bar T$ deformation,''
Nucl. Phys. B {\bf 958}, 115119 (2020)
doi:10.1016/j.nuclphysb.2020.115119
[arXiv:2002.04824 [hep-th]].
}

\lref\Chang{
C.~K.~Chang, C.~Ferko and S.~Sethi,
``Supersymmetry and $ T\overline{T} $ deformations,''
JHEP {\bf 04}, 131 (2019)
doi:10.1007/JHEP04(2019)131
[arXiv:1811.01895 [hep-th]].
}

\lref\Drukker{
N.~Drukker, B.~Fiol and J.~Simon,
``Godel's universe in a supertube shroud,''
Phys. Rev. Lett. {\bf 91}, 231601 (2003)
doi:10.1103/PhysRevLett.91.231601
[arXiv:hep-th/0306057 [hep-th]].
}

\lref\Gimon{
E.~G.~Gimon and P.~Horava,
``Over-rotating black holes, Godel holography and the hypertube,''
[arXiv:hep-th/0405019 [hep-th]].
}

\lref\Baggio{
M.~Baggio, A.~Sfondrini, G.~Tartaglino-Mazzucchelli and H.~Walsh,
``On $ T\overline{T} $ deformations and supersymmetry,''
JHEP {\bf 06}, 063 (2019)
doi:10.1007/JHEP06(2019)063
[arXiv:1811.00533 [hep-th]].
}

\lref\Forste{
S.~Forste,
``A Truly marginal deformation of SL(2, R) in a null direction,''
Phys. Lett. B {\bf 338}, 36-39 (1994)
doi:10.1016/0370-2693(94)91340-4
[arXiv:hep-th/9407198 [hep-th]].
}

\lref\IsraelM{
D.~Israel, C.~Kounnas and M.~P.~Petropoulos,
``Superstrings on NS5 backgrounds, deformed AdS(3) and holography,''
JHEP {\bf 10}, 028 (2003)
doi:10.1088/1126-6708/2003/10/028
[arXiv:hep-th/0306053 [hep-th]].
}

\lref\IsraelMM{
D.~Israel,
``Quantization of heterotic strings in a Godel / anti-de Sitter space-time and chronology protection,''
JHEP {\bf 01}, 042 (2004)
doi:10.1088/1126-6708/2004/01/042
[arXiv:hep-th/0310158 [hep-th]].
}

\lref\KiritsisM{
E.~Kiritsis,
``Exact duality symmetries in CFT and string theory,''
Nucl. Phys. B {\bf 405}, 109-142 (1993)
doi:10.1016/0550-3213(93)90428-R
[arXiv:hep-th/9302033 [hep-th]].
}

\lref\HorowitzM{
G.~T.~Horowitz and A.~A.~Tseytlin,
``A New class of exact solutions in string theory,''
Phys. Rev. D {\bf 51}, 2896-2917 (1995)
doi:10.1103/PhysRevD.51.2896
[arXiv:hep-th/9409021 [hep-th]].
}

\lref\Sadri{
D.~Sadri and M.~M.~Sheikh-Jabbari,
``String theory on parallelizable pp waves,''
JHEP {\bf 06}, 005 (2003)
doi:10.1088/1126-6708/2003/06/005
[arXiv:hep-th/0304169 [hep-th]].
}

\lref\Frolov{
S.~Frolov,
``Lax pair for strings in Lunin-Maldacena background,''
JHEP {\bf 05}, 069 (2005)
doi:10.1088/1126-6708/2005/05/069
[arXiv:hep-th/0503201 [hep-th]].
}

\lref\Mateos{
D.~Mateos and P.~K.~Townsend,
``Supertubes,''
Phys. Rev. Lett. {\bf 87}, 011602 (2001)
doi:10.1103/PhysRevLett.87.011602
[arXiv:hep-th/0103030 [hep-th]].
}

\lref\Bousso{
R.~Bousso,
``The Holographic principle for general backgrounds,''
Class. Quant. Grav. {\bf 17}, 997-1005 (2000)
doi:10.1088/0264-9381/17/5/309
[arXiv:hep-th/9911002 [hep-th]].
}

\lref\HawkingM{
S.~W.~Hawking,
``The Chronology protection conjecture,''
Phys. Rev. D {\bf 46}, 603-611 (1992)
doi:10.1103/PhysRevD.46.603
}

\lref\Johnson{
C.~V.~Johnson, A.~W.~Peet and J.~Polchinski,
``Gauge theory and the excision of repulson singularities,''
Phys. Rev. D {\bf 61}, 086001 (2000)
doi:10.1103/PhysRevD.61.086001
[arXiv:hep-th/9911161 [hep-th]].
}

\lref\SomMM{
M.~M.~Som and  A.~K.~Raychaudhuri,
``Cylindrically symmetric charged dust distributions in rigid rotation in general relativity,"
Proc. R. Soc. Lond. A {\bf 304} 81-86 (1968)
doi:10.1098/rspa.1968.0073
}

\lref\Brecher{
D.~Brecher, P.~A.~DeBoer, D.~C.~Page and M.~Rozali,
``Closed time - like curves and holography in compact plane waves,''
JHEP {\bf 10}, 031 (2003)
doi:10.1088/1126-6708/2003/10/031
[arXiv:hep-th/0306190 [hep-th]].
}

\lref\WilsonMM{
K.~G.~Wilson and W.~Zimmermann,
``Operator product expansions and composite field operators in the general framework of quantum field theory,''
Commun. Math. Phys. {\bf 24} (1972), 87-106
doi:10.1007/BF01878448
}

\lref\WittenM{
E.~Witten,
``Multitrace operators, boundary conditions, and AdS / CFT correspondence,''
[arXiv:hep-th/0112258 [hep-th]].
}

\lref\AldayMM{
L.~F.~Alday, G.~Arutyunov and S.~Frolov,
``Green-Schwarz strings in TsT-transformed backgrounds,''
JHEP {\bf 06}, 018 (2006)
doi:10.1088/1126-6708/2006/06/018
[arXiv:hep-th/0512253 [hep-th]].
}

\lref\Lowenstein{
J.~H.~Lowenstein,
``Normal products in the thirring model,''
Commun. Math. Phys. {\bf 16}, 265-289 (1970)
doi:10.1007/BF01646535
}

\lref\ApoloMM{
L.~Apolo, P.~X.~Hao, W.~X.~Lai and W.~Song,
``Glue-on AdS holography for $ T\overline{T} $-deformed CFTs,''
JHEP {\bf 06}, 117 (2023)
doi:10.1007/JHEP06(2023)117
[arXiv:2303.04836 [hep-th]].
}

\lref\ManschotMM{
J.~Manschot and S.~Mondal,
``Supersymmetric black holes and $T{\bar T}$ deformation,''
Phys. Rev. D {\bf 107}, no.12, L121903 (2023)
doi:10.1103/PhysRevD.107.L121903
[arXiv:2207.01462 [hep-th]].
}

\lref\MaldacenaMM{J.~M.~Maldacena,
``The Large N limit of superconformal field theories and supergravity,''
Adv. Theor. Math. Phys. {\bf 2}, 231-252 (1998)
doi:10.4310/ATMP.1998.v2.n2.a1
[arXiv:hep-th/9711200 [hep-th]].
}

\lref\GubserMM{
S.~S.~Gubser, I.~R.~Klebanov and A.~M.~Polyakov,
``Gauge theory correlators from noncritical string theory,''
Phys. Lett. B {\bf 428}, 105-114 (1998)
doi:10.1016/S0370-2693(98)00377-3
[arXiv:hep-th/9802109 [hep-th]].
}

\lref\WittenMM{
E.~Witten,
``Anti-de Sitter space and holography,''
Adv. Theor. Math. Phys. {\bf 2}, 253-291 (1998)
doi:10.4310/ATMP.1998.v2.n2.a2
[arXiv:hep-th/9802150 [hep-th]].
}

\lref\MMRazamat{
O.~Aharony, Z.~Komargodski and S.~S.~Razamat,
``On the worldsheet theories of strings dual to free large N gauge theories,''
JHEP {\bf 05}, 016 (2006)
doi:10.1088/1126-6708/2006/05/016
[arXiv:hep-th/0602226 [hep-th]].
}

\lref\HerdeiroMM{
C.~A.~R.~Herdeiro,
``Spinning deformations of the D1 - D5 system and a geometric resolution of closed timelike curves,''
Nucl. Phys. B {\bf 665}, 189-210 (2003)
doi:10.1016/S0550-3213(03)00484-X
[arXiv:hep-th/0212002 [hep-th]].
}

\lref\KutasovME{
D.~Kutasov and N.~Seiberg,
``More comments on string theory on AdS(3),''
JHEP {\bf 04}, 008 (1999)
doi:10.1088/1126-6708/1999/04/008
[arXiv:hep-th/9903219 [hep-th]].
}

\lref\AharonyMMR{
O.~Aharony, M.~Berkooz and E.~Silverstein,
``Multiple trace operators and nonlocal string theories,''
JHEP {\bf 08}, 006 (2001)
doi:10.1088/1126-6708/2001/08/006
[arXiv:hep-th/0105309 [hep-th]].
}

\lref\BeckerME{
K.~Becker and M.~Becker,
``Interactions in the SL(2,IR) / U(1) black hole background,''
Nucl. Phys. B {\bf 418}, 206-230 (1994)
doi:10.1016/0550-3213(94)90245-3
[arXiv:hep-th/9310046 [hep-th]].
}

\lref\FDavidMM{
F.~David,
``Conformal Field Theories Coupled to 2D Gravity in the Conformal Gauge,''
Mod. Phys. Lett. A {\bf 3}, 1651 (1988)
doi:10.1142/S0217732388001975
}

\lref\ChaudhuriS{
S.~Chaudhuri and J.~A.~Schwartz,
``A Criterion for Integrably Marginal Operators,''
Phys. Lett. B {\bf 219}, 291-296 (1989)
doi:10.1016/0370-2693(89)90393-6
}

\lref\Courant{
T.~Courant,
``Dirac manifolds,''
Trans.Am.Math.Soc. {\bf 319} (1990) 631
doi:10.1090/S0002-9947-1990-0998124-1
}

\lref\HenneauxMMT{
M.~Henneaux, C.~Martinez, R.~Troncoso and J.~Zanelli,
``Asymptotically anti-de Sitter spacetimes and scalar fields with a logarithmic branch,''
Phys. Rev. D {\bf 70}, 044034 (2004)
doi:10.1103/PhysRevD.70.044034
[arXiv:hep-th/0404236 [hep-th]].
}

\lref\ReggeMMT{
T.~Regge and C.~Teitelboim,
``Role of Surface Integrals in the Hamiltonian Formulation of General Relativity,''
Annals Phys. {\bf 88}, 286 (1974)
doi:10.1016/0003-4916(74)90404-7
}

\lref\HenneauxMMTT{
M.~Henneaux, C.~Martinez, R.~Troncoso and J.~Zanelli,
``Asymptotic behavior and Hamiltonian analysis of anti-de Sitter gravity coupled to scalar fields,''
Annals Phys. {\bf 322}, 824-848 (2007)
doi:10.1016/j.aop.2006.05.002
[arXiv:hep-th/0603185 [hep-th]].
}

\lref\Sokoloff{
D.~D.~Sokoloff, A.~A.~Starobinskii, ``On the structure of curvature tensor on conical singularities,"
Dokl. Akad. Nauk SSSR, 1977, Volume {\bf 234}, Number 5, 1043-1046
}

\lref\Kowalski{
Kowalski, O.~, Szenthe, J~. ``On the Existence of Homogeneous Geodesics in Homogeneous Riemannian Manifolds,"
Geometriae Dedicata {\bf 81}, 209-214 (2000)
doi:10.1023/A:1005287907806
}

\lref\GauntlettNM{
J.~P.~Gauntlett, J.~B.~Gutowski, C.~M.~Hull, S.~Pakis and H.~S.~Reall,
``All supersymmetric solutions of minimal supergravity in five- dimensions,''
Class. Quant. Grav. {\bf 20}, 4587-4634 (2003)
doi:10.1088/0264-9381/20/21/005
[arXiv:hep-th/0209114 [hep-th]].
}

\lref\MaldacenaDLC{
J.~Maldacena, D.~Martelli and Y.~Tachikawa,
``Comments on string theory backgrounds with non-relativistic conformal symmetry,''
JHEP {\bf 10}, 072 (2008)
doi:10.1088/1126-6708/2008/10/072
[arXiv:0807.1100 [hep-th]].
}

\lref\CallanCGS{
C.~G.~Callan, Jr., E.~J.~Martinec, M.~J.~Perry and D.~Friedan,
``Strings in Background Fields,''
Nucl. Phys. B {\bf 262}, 593-609 (1985)
doi:10.1016/0550-3213(85)90506-1
}

\lref\DijkgraafMM{
R.~Dijkgraaf, H.~L.~Verlinde and E.~P.~Verlinde,
``String propagation in a black hole geometry,''
Nucl. Phys. B {\bf 371}, 269-314 (1992)
doi:10.1016/0550-3213(92)90237-6
}

\lref\GinspargM{
P.~H.~Ginsparg and F.~Quevedo,
``Strings on curved space-times: Black holes, torsion, and duality,''
Nucl. Phys. B {\bf 385}, 527-557 (1992)
doi:10.1016/0550-3213(92)90057-I
[arXiv:hep-th/9202092 [hep-th]].
}

\lref\SfetsosM{
K.~Sfetsos,
``Conformally exact results for SL(2,R) x SO(1,1)(d-2) / SO(1,1) coset models,''
Nucl. Phys. B {\bf 389}, 424-444 (1993)
doi:10.1016/0550-3213(93)90327-L
[arXiv:hep-th/9206048 [hep-th]].
}

\lref\Codina{
T.~Codina, O.~Hohm and B.~Zwiebach,
``On black hole singularity resolution in $D=2$ via duality-invariant $\alpha'$ corrections,''
[arXiv:2308.09743 [hep-th]].
}

\lref\WittenMMM{
E.~Witten,
``On string theory and black holes,''
Phys. Rev. D {\bf 44}, 314-324 (1991)
doi:10.1103/PhysRevD.44.314
}

\lref\Mandal{
G.~Mandal, A.~M.~Sengupta and S.~R.~Wadia,
``Classical solutions of two-dimensional string theory,''
Mod. Phys. Lett. A {\bf 6}, 1685-1692 (1991)
doi:10.1142/S0217732391001822
}

\lref\GiveonMMMMMM{
A.~Giveon,
``Target space duality and stringy black holes,''
Mod. Phys. Lett. A {\bf 6}, 2843-2854 (1991)
doi:10.1142/S0217732391003316
}

\lref\HullMMMM{
C.~M.~Hull,
``Timelike T duality, de Sitter space, large N gauge theories and topological field theory,''
JHEP {\bf 07}, 021 (1998)
doi:10.1088/1126-6708/1998/07/021
[arXiv:hep-th/9806146 [hep-th]].
}

\Title{
}
{\vbox{
\centerline{Moving holographic boundaries} }}

\bigskip
\centerline{\it Meseret Asrat}
\smallskip
\centerline{International Center for Theoretical Sciences}
\centerline{Tata Institute of Fundamental Research
} \centerline{Bengaluru, KA 560089, India}

\smallskip

\vglue .3cm

\bigskip

\let\includefigures=\iftrue
\bigskip
\noindent

In this paper, we show that for one sign of the deformation coupling single-trace $T{\bar T}$ deformation moves the holographic screen in G\"{o}del universe radially inward. For the other sign of the coupling it moves the holographic screen radially outward. We (thus) argue, on general grounds, that in holography (single-trace) $T{\bar T}$ deformation can be generally thought of as either moving the holographic boundary into the bulk or washing it away to infinity. In Anti-de Sitter this breaks the spacetime conformal symmetry. We further note that moving timelike holographic boundary into bulk creates a curvature singularity. In the boundary the singularity is understood by states with imaginary energies. To make the theory sensible we introduce an ultraviolet cutoff and thereby move the boundary into the bulk. In this paper we first obtain the Penrose limit of the single-trace $T{\bar T}$ deformed string background and then perform T-duality along a space-like isometry to obtain a class of deformed G\"{o}del universes. The string background we consider is $AdS_3\times S^3\times {\cal M}_4$. The single-trace $T{\bar T}$ deformation is a particular example of the more general $O(d, d)$ transformations. 

\bigskip

\Date{04/23}

\newsec{Introduction}

In recent years, there have been two independent but related developments in the study of irrelevant deformations in two dimensional quantum field theories (QFTs). In these developments the deformations are generated by special operators \refs{\Zamolodchikov\Smirnov\Cavaglia-\Giveon}. The deformations do not mix states, \ie, a given initial state gets deformed to a new state which depends only on the initial given state (and the deformation coupling). The deformations also do not create or introduce new states. Thus, different sectors of a Hilbert space deform independently in a one-to-one fashion \refs{\Aharony\AsratT-\Asrat}. In general, depending on the sign of the deformations couplings, the energy spectrum of the deformed theory (quantized on a circle) is either real or complex. We will comment on this later in the paper.

The first development is the deformation of a two dimensional QFT by the determinant of the energy momentum stress tensor, commonly referred to as $T{\bar T}$ deformation \refs{\Zamolodchikov\Smirnov-\Cavaglia}. The generating $T{\bar T}$ operator is irrelevant (in the Wilsonian renormalization group flow sense\foot{Note irrelevance is defined with respect to a fixed point. An irrelevant operator is less important in the infrared but become more important in the ultraviolet.}) and it is well-defined (in the coincidence limit up to total derivative terms\foot{Operator product expansion coefficients are naturally interpreted as renormalization counter terms \refs{\WilsonMM,\ \Lowenstein}. Thus, $T{\bar T}$ (is expected to) results in a theory which is well-behaved at high energies, \ie, under a group of scale transformations or redefinition of the the cut-off sclae.}) \Zamolodchikov. Since it is build out of (the components of) the stress tensor which is present and conserved in any QFT with spacetime translation invariance, the operator exists universally, and therefore also the deformation is universally applicable. Given that the operator is well-defined in any QFT, it can be used to define a one parameter family of theories by iteratively adding it to and updating the Lagrangian density in a small increment of the deformation coupling. This defines a curve or trajectory in field theory space (parametrized by the coupling). Therefore, it can be used to probe and better understand the space of field theories.

In general, under an irrelevant deformation of a conformal field theory (CFT), the understanding is that at high energies or in the ultraviolet (UV) the (quantum) theory is ill-defined. The theory involves ambiguities and/or singularities and thus, the deformation is not solvable. In general we may, but not necessarily, attribute this to the existence of infinitely many irrelevant operators at the fixed point. See \Asrat\ for further discussion.

The $T{\bar T}$ deformation, however, defines or results in a theory that is under better mathematical tractability. In particular, the deformation preserves some of the symmetries of the original theory, and in the resulting deformed theory, for several quantities including quantum observables (\eg, Lagrangian density, currents, \etc.), exact and finite or non-singular results are obtained. Among these, the principal quantity of interest is the energy (or the canonical partition function). Thus, in this sense the deformation is solvable.

The energy spectrum (and hence the canonical partition function) of the deformed theory (quantized on a circle\foot{There is a minimum size of the circle below which the Casimir energy or the effective central charge becomes complex, see \Asrat\ and references therein.}) is obtained (in particular in the case the original theory is a CFT) using the operator product expansion (OPE) technique \WilsonMM. See also equation (C.7) in Appendix C. At short distances or high energies the deformed theory exhibits Hagedorn density of states and non-locality. Thus, at high energies the deformed theory is not governed by a conventional local fixed point, see \Asrat\ and references therein.

In the case the undeformed theory is a holographic CFT \ie, dual to string theory in $AdS_3$ (times some compact internal space), the dual spacetime after $T{\bar T}$ deformation\foot{For one sign of the coupling. In our notation the sign is positive. See also \refs{\ApoloMM,\ \ManschotMM} for discussions in the case the sign of the coupling is negative.} is proposed to be $AdS_3$ with a Dirichlet boundary condition at a finite radial distance which is fixed by the deformation coupling. Thus, in this sense, $T{\bar T}$ deformation is proposed to move the boundary where the CFT is thought to live into the bulk \McGough.

The second development is in two dimensional holographic field theories which are dual to string theory in asymptotically Anti-de Sitter (AdS) spacetimes \Giveon. In this latter development, the deformation is commonly referred to as single-trace $T{\bar T}$ deformation for a reason that we will explain momentarily. It is also irrelevant at the infrared (IR) fixed point. Unlike the usual $T{\bar T}$ deformation (discussed above), however, as we will explain shortly, it is defined from the bulk (worldsheet) string theory side. Thus, it is only (or it is at least) applicable or defined in two dimensional holographic CFTs. This deformation is closely related but distinct from the usual $T{\bar T}$ deformation (discussed above).

In $AdS_3$ the long string sector is believed to be described at weak coupling or near the boundary and large $N$ (where $N$ is the maximal number of long strings) by a symmetric product CFT. Therefore, in this sector the single-trace $T{\bar T}$ operator is (believed to be) given by sum of (the usual) $T{\bar T}$ operators and therefore the adjective single-trace.\foot{Its existence is related to a (symmetric) product structure of the (dual) boundary theory.} In general in the case the holographic theory is described by a symmetric product, the single-trace $T{\bar T}$ operator is (believed to be) given by a sum of (the usual) $T{\bar T}$ operators. The sum is over the copies of the defining theory of the symmetric product. In this case the usual $T{\bar T}$ operator is given by product of two sums. The two factors in the product are the two components of the total energy momentum tensor of the symmetric product CFT. Thus, the usual $T{\bar T}$ deformation is also sometimes referred to as double-trace $T{\bar T}$ deformation.\foot{See \refs{\WittenM, \ \AharonyMMR} for a discussion on multi-trace operators and holography.} 

In general, irrespective of whether the boundary theory (which is dual to the (combined) short and long string sectors) is described by a symmetric product or not, in the bulk, the deformation is understood as a T-duality--shift--T-duality (TsT) transformation\foot{Such TsT transformations can be equivalently viewed as twisted boundary conditions, see \Asrat\ and references therein. This will become more evident in section two. In this view the background spacetime remains locally $AdS_3$. Thus, the local dynamics of strings in the deformed background is described by the $SL(2, R)$ Wess-Zumino-Witten (WZW) sigma model at level $k$. This observation was noted and used to compute spectrum and correlation functions in \Asrat.} or as a particular example of the more general $O(d,d)$ transformations \refs{\Lunin,\ \Frolov, \ \IsraelM\AldayMM\Araujo\ApoloTT-\Sfondrini}. $O(d,d)$ transformations are used to generate new string theory backgrounds from known ones. They correspond to deforming the worldsheet theory by a linear combination of exactly marginal current bilinear operators \Hassan.

Note that the identification of the bulk marginal current bilinear deformation with the irrelevant single-trace boundary deformation is at the level of the respective theories actions \Giveon. This is crucial for the identification to make sense. We explain this below.

In general, in classical (super)gravity (or $1/N$ expansion) a bulk operator $\phi(x, \rho)$ whose asymptotic behavior is $\rho^{-\Delta}$ is identified via the gauge/(super)gravity duality \refs{\MaldacenaMM\GubserMM-\WittenMM} with a boundary operator ${\cal O}(x) \equiv \lim_{\rho\to \infty}\rho^{\Delta}\phi(x, \rho)$ whose (scaling) dimension is $\Delta$, where $x$ denotes collectively the boundary spacetime coordinates.\foot{A scalar $\phi$ of mass $m$ in $AdS_{d + 1}$ has in general the asymptotic expansion of the form $\phi \approx {\rm source\ term}\cdot \rho^{-(d - \Delta)} + {\rm expectation\ value }\cdot \rho^{-\Delta}$ where $\Delta = d/2 + \sqrt{d^2/4 + m^2l^2}$, and $l$ is the radius of $AdS_{d + 1}$. See \refs{\ReggeMMT\HenneauxMMT-\HenneauxMMTT} for special cases and more discussion.} Of the (several) classes of operators ${\cal O}(x)$, the class (consisting) of single trace operators is of particular interest. The (gauge invariant) single trace boundary operators ${\cal O}(x)$ can be equivalently given as integrated (physical) vertex operators ${\cal V}(z, x)$ on the worldsheet, \ie, ${\cal O}(x) \equiv \int d^2z{\cal V}(z, x)$. Therefore, one can write $\int d^2x{\cal O}(x) \equiv \int d^2z{\cal K}(z)$, where we defined ${\cal K}(z) :=\int d^2x {\cal V}(x, z)$.\foot{In general, however we cannot exchange order of integrations.} Adding $\int d^2z{\cal K}(z)$ to the worldsheet action or equivalently ${\cal K}(z)$ to the worldsheet Lagrangian is then (expected to be) equivalent via the duality to adding $\int d^2x{\cal O}(x)$ to the boundary theory action or equivalently ${\cal O}(x)$ to the boundary theory Lagrangian. It is in this sense the equivalence or identification is made in \Giveon. Note that ${\cal K}(z)$ is obtained by integrating ${\cal V}(x, z)$ over all the boundary spacetime. 

The (gauge invariant) single trace operator ${\cal O}(x)$ that is considered in \Giveon\ is studied earlier in \KutasovME. We follow the notations of \refs{\Giveon, \ \KutasovME} and denote the operator by $D(x)$, see Appendix C for its expression. The corresponding operator ${\cal K}(z)$ is a current bilinear and exactly marginal \Giveon. It is (related to) the screening charge at strictly large level (of the $sl(2)$ affine (current) Lie algebra), \ie, $\int d^2 z {\cal K} = \int d^2 z \beta{\bar \beta}$, see \eg, \BeckerME.\foot{The (relevant) screening charge is given by $Q_k = \int d^2z\beta\bar\beta e^{-{2\over \alpha_+}\phi}$, where $\alpha_+ = \sqrt{2(k - 2)}$ and $k$ is the level. The operator $e^{-{2\over \alpha_+}\phi}$ has scaling dimensions zero.} $D(x)$ is the dilaton vertex operator and has the same quantum numbers as the composite operator $T{\bar T}(x)$ \KutasovME. This is one of the primary reasons for identifying $D(x)$ with a/the single trace $T{\bar T}(x)$. However, $D(x)$ is different from (the canonical) $T{\bar T}(x)$ since the latter is given (via the duality) by a product of two integrated (physical) vertex operators corresponding to $T$ and ${\bar T}$.\foot{If we write $T(x) =\int d^2z {\cal V}_{T}(z, x)$ and ${\bar T}(y) = \int d^2z {\cal V}_{\bar T}(z, y)$, then the usual operator $T{\bar T}(x)$ is given by $T{\bar T}(x) = \lim_{y\to x}\int d^2z\int d^2w{\cal V}_{T}(z, x){\cal V}_{\bar T}(w, y)$. This implies $\int d^2xT{\bar T}(x) = \int d^2z {\cal M}_{T\bar T}(z)$ provided ${\cal M}_{T\bar T}(z)$ exists. The operator ${\cal M}_{T\bar T}(z) := \lim_{\epsilon\to 0}\int d^2 w\int d^2x {\cal V}_{T}(z, x){\cal V}_{\bar T}(w, x + \epsilon)$. In view of the $T{\bar T}$ operator, we expect ${\cal M}_{T\bar T}(z)$ to exist and be well-defined although it is obtained by integrating over all worldsheet and spacetime coordinates, see \refs{\AharonyMMR, \ \MMRazamat} for related discussions.} However, in the (long string) untwisted sector $D(x)$ should agree with the usual $T{\bar T}(x)$ (of the defining theory of the orbifold theory), assuming the identification is correct. As we mentioned earlier, the existence of $D(x)$ is related to the orbifold structure of the boundary theory.

The single-trace $T{\bar T}$ deformation, for one sign of the coupling \foot{In our notation the sign is negative.} and in Poincar\'e coordinates, corresponds in the bulk to a string background that interpolates between $AdS_3 \times {\cal M}_7$ in the IR and a linear dilaton spacetime (or vacuum of compactified little string theory (LST)) $\IR^{(1,1)}\times \IR  \times {\cal M}_7$ in the UV, where ${\cal M}_7$ is a compact seven dimensional (internal) manifold. In this paper, we determine the deformed string background in global coordinates and discuss both signs. To the best of my knowledge, this result has not been obtained and reported elsewhere in the literature, and one of the main goals of this paper is to present and discuss this result.

The single-trace $T{\bar T}$ deformation is solvable in the sense that (in the long string sector) the exact energy spectrum can be obtained. See equation (C.7) in Appendix C. At high energies the deformed theory also exhibits Hagedorn density of states and non-locality, see \Asrat\ and references therein.

In this paper we study G\"odel (type) universes \refs{\Godel\SomMM-\Gauntlett} in the single-trace $T{\bar T}$ deformed string theory, or more generally, in string theory obtained by $O(d,d)$ transformations \Araujo. We use G\"odel (type) universe (as a toy model) to shed light on (single-trace) $T{\bar T}$ deformation in holography. As a consequence of (the) factorization property (of the $T{\bar T}$ operator), the $T{\bar T}$ deformation does not mix different sectors of a Hilbert space and also does not create or introduce new states. This is also the case in single-trace $T{\bar T}$ deformation (in particular in the long string sector), see \eg, \Asrat\ and references therein. In string theory G\"odel (type) universes are (usually) obtained following these two steps: (1) we take the Penrose limit, \ie\ restrict the Hilbert space to a specific sector and (2) we perform T-duality along a spacelike isometry \refs{\Boyda-\Harmark}. See also \Brecher.

The G\"odel (type) universe is characterized by a preferred (cylindrical) surface, which is referred to as a holographic screen \refs{\Boyda, \ \Bousso}\foot{In the rest of the paper we will use this terminology to refer to the preferred surface.}, and closed timelike curves. The closed timelike curves do not constitute geodesics and (must) intersect the region beyond the holographic screen \Hawking. Therefore, the holographic screen, as defined by an inertial observer, either hides \ie, screens, the closed timelike curves or cuts them into parts. Thus, in classical (super)gravity it prevents the development of closed timelike curves.

It is conjectured in \refs{\Boyda} that in general holography protects causality \refs{\HawkingM}.  A different view is proposed in the paper \Drukker\ by studying (cylindrical) supertube probes \Mateos. It shares the same underlying concept with the enhan\c{c}on approach \Johnson, but also has some (contextual) differences. It replaces the spacetime beyond the holographic screen by a spacetime which is free of closed timelike curves (and singularities). However, its status is not clear, see \eg, \refs{\Brace,\ \Gimon}.

In this paper we will not be concerned on understanding the closed timelike curves. They do not directly play any role and also they do not pose any problem in our discussion. We will leave that to a future work. For discussions on quantization of strings in G\"odel (type) universe see also \refs{\Brace, \ \IsraelMM}.

In this paper we show that, depending on the sign of the deformation coupling, single-trace $T{\bar T}$ deformation corresponds to moving the holographic screen either into the bulk or out to infinity. Therefore, in holography in general it is natural to view the single-trace $T{\bar T}$ deformation or TsT\foot{The shift is in the time coordinate.} as moving the boundary into the bulk or washing it away to infinity. We explain in what sense. This observation is one of the other several main results of the paper. Note that as we explained earlier the single and double trace $T{\bar T}$ deformations are in general different. However, our result and \McGough\ provide another evidence that the deformations are closely related. We will comment on the main difference of the proposal \McGough\ and our result in section six.

The paper is organized as follows. In section two we obtain the single-trace $T{\bar T}$ deformed string theory background in global coordinates by performing TsT transformation on $AdS_3\times S^3\times {\cal M}_4$. We also show that in the Poincar\'e (patch) limit it reduces to the result given in \refs{\Giveon, \ \IsraelM,\ \Forste}. In section three we briefly review the Penrose limit. Specifically, we obtain a Penrose limit of the deformed string theory. In section four we perform T-duality along a spacelike isometry and get the single-trace $T{\bar T}$ deformed G\"odel (type) universe. We discuss how the location of the holographic screen depends on the coupling of the deformation. We also discuss the implications for (single-trace $T{\bar T}$ deformation and) holography in section five. In section six we discuss future research directions. 

In the appendices we review and recall some (known) results that we find relevant to make our discussion as clear as possible. In Appendix A we review the T-duality transformation in type II bosonic string theory.  In Appendix B we review the (single trace) $D$ operator which is identified in the long string sector with the single trace $T{\bar T}$ operator. In Appendix C we discuss the deformation in the worldsheet and give the deformed boundary spacetime spectrum. In Appendix D we provide the Maurer-Cartan one form on the three sphere. We use the one form in section four. In Appendix E we give the fundamental domain of the deformation coupling.

\newsec{The $T{\bar T}$ deformed $AdS_3$ string background}

We consider string theory on $AdS_3\times S^3 \times {\cal M}_4$ where ${\cal M}_4$ is a compact four dimensional manifold. The simplest examples are the Calabi-Yau manifolds $T^4$ (which breaks no supercharges) and $K_3$ (which breaks half of the supercharges). The background in general supports admixture of Neveu-Schwartz--Neveu-Schwartz (NSNS) and Ramond-Ramond (RR) fluxes. In this paper we only consider bosonic string theory.

The $AdS_3$ part of the worldsheet string action in the presence of NSNS flux is given by the Wess-Zumino-Witten (WZW) sigma model on the group manifold $SL(2, R)$\foot{In general superstring on $AdS_3\times S^3$ supported with NSNS and RR fluxes is described by the sigma model on the supergroup $PSU(1, 1|2) \cong SU(1,1|2)/C^*$ where $C^*$ is the multiplicative group of complex numbers \refs{\Berkovits,\ \Gotz}. $C^*$ is isomorphic to the product $R^*_+\times U(1)$ where $R^*_+$ is the multiplicative group of positive real numbers. In general the bosons (even subspace) and fermions (odd subspace) are coupled (except, \eg\ with no RR flux, at strictly large level $k$ and/or $N$). However, the single-trace $T{\bar T}$ deformation acts only on the $SU(1, 1) \cong SL(2, R)$ part of the even subspace. It doesn't alter the interaction term, \eg, in the case the number of RR fluxes is zero and $k$ is finite.}
\eqn\aaaa{S^{SL(2,R)}_k = {k\over 2 \pi}\int d^2z \left[\partial\tilde\theta\bar\partial\tilde\theta - \cosh^2\tilde\theta \partial\tilde\varphi\bar\partial\tilde\varphi + \sinh^2\tilde\theta\partial\tilde\psi\bar\partial\tilde\psi - {1\over 2}\cosh(2\tilde\theta)\left(\bar\partial\tilde\varphi\partial\tilde\psi - \bar\partial\tilde\psi\partial\tilde\varphi\right)\right],  
}
where $k$ is the level. The coordinate $\tilde\psi$ is an angle variable and it has $2\pi$ periodicity.

The $S^3$ part of the worldsheet string action is given by the WZW sigma model on the group manifold $SU(2)$,
\eqn\stringonsphere{S_k^{SU(2)} = {k\over 2\pi}\int d^2z \left[\partial\theta\bar\partial\theta + \cos^2\theta \partial \varphi\bar\partial\varphi + \sin^2\theta\partial\psi\bar\partial\psi + {1\over2}\cos(2\theta)\left(\bar\partial\varphi \partial\psi - \bar\partial\psi\partial\varphi\right)\right].
}
See Appendix D for details of the parametrization of the group $SU(2)$ we are using.

The single-trace $T{\bar T}$ deformed string action is obtained as follows. We will ignore the $S^3\times {\cal M}_4$ part of the action for a moment. We first perform T-duality along $\tilde\psi$ \refs{\Buscher\BuscherB-\Rocek,\ \Alvarez, \ \BerkovitsB}. Along the string, the angular variable $\tilde\psi(\tau, \sigma + 2\pi) \sim \tilde\psi(\tau, \sigma) + 2\pi w$, where $w$ is some integer. See Appendix A for details on how the background fields and action transform under the T-duality. This gives
\eqn\aabb{\eqalign{S_k = 
 {k\over 2 \pi}\int d^2z \left[\partial\tilde\theta\bar\partial\tilde\theta + \left(- \cosh^2\tilde\theta  + {1\over 4}{\cosh^22\tilde\theta\over \sinh^2\tilde\theta}\right)\partial\tilde\varphi\bar\partial\tilde\varphi + {1\over \sinh^2\tilde\theta}\partial\tilde\chi\bar\partial\tilde\chi\right.\cr 
 - \left.{1\over 2}{\cosh(2\tilde\theta)\over \sinh^2\tilde\theta}(\bar\partial\tilde\varphi\partial\tilde\chi + \bar\partial\tilde\chi\partial\tilde\varphi)\right]. 
}
}
Note that there is no WZW term in the above action. Also, at constant $\tilde\varphi$, there is no conical singularity at $\tilde\theta = \infty$ (and of course also at $\tilde\theta = 0$). At $\tilde\theta = 0$ the Ricci invariant is divergent. It is given by $R = -4/\sinh^2\tilde\theta$. Thus, in general $\tilde\chi$ need not be periodic with a definite period to avoid a conical singualrity. The period of $\tilde\psi$ in \aaaa\ is chosen to be $2\pi$ in order to avoid conical singularity at $\tilde\theta = 0$.\foot{The period of a dual coordinate, in general, is fixed indirectly by the duality. The duality transformation in general maps a circle of radius $r$ to a dual circle of radius $\alpha'/r$. Therefore, the period of a coordinate on the dual circle is $2\pi \alpha' /r$, \eg, see \DijkgraafMM.}

We next shift the coordinate $\tilde\varphi$ as
\eqn\aacc{\tilde\varphi \to \tilde\varphi + {1\over 2}\lambda \tilde\chi.
}
Note this is different from the shift in the usual TsT transformation. The shift in the usual TsT transformation is in a spatial direction. The parameter $\lambda$ is dimensionless. It is related to the (single-trace) $T{\bar T}$ deformation coupling $\hat\lambda$ by $\hat\lambda = \lambda l_s^2$ (up to a constant factor) where $l_s = \sqrt{\alpha'}$ is the string length. The action now becomes 
\eqn\aadd{\eqalign{S_k = 
{k\over 2 \pi} \int d^2z \left\{\partial\tilde\theta\bar\partial\tilde\theta + \left(- \cosh^2\tilde\theta  + {1\over 4}{\cosh^22\tilde\theta\over \sinh^2\tilde\theta}\right)\partial\tilde\varphi\bar\partial\tilde\varphi \right.\cr 
+ \left[- {1\over 2}{\cosh(2\tilde\theta)\over \sinh^2\tilde\theta} +  {\lambda\over 2} \left(- \cosh^2\tilde\theta  + {1\over 4}{\cosh^22\tilde\theta\over \sinh^2\tilde\theta}\right)\right]\left(\bar\partial\tilde\varphi\partial\tilde\chi + \bar\partial\tilde\chi\partial\tilde\varphi\right) \cr
 +\left.\left[{1\over \sinh^2\tilde\theta}+ {\lambda\over 2} \left(-{\cosh(2\tilde\theta)\over \sinh^2\tilde\theta} +  {\lambda\over2} \left(- \cosh^2\tilde\theta  + {1\over 4}{\cosh^22\tilde\theta\over \sinh^2\tilde\theta}\right)\right)\right]\bar\partial\tilde\chi\partial\tilde\chi\right\}.
}
}
Some remarks are in order now.

The coefficient of $\bar\partial\tilde\chi\partial\tilde\chi$ is positive for negative $\lambda$. For positive $\lambda$ it can either be positive or negative. It is zero at $\tilde\theta = -{1\over 2}\ln \left({\lambda\over 4}\right) := \tilde\theta_0$. In what follows we assume the coefficient of $\bar\partial\tilde\chi\partial\tilde\chi$ is positive. Therefore, this restricts the allowed values of the coupling to the region $\lambda \leq 4$. At $\lambda = 4$, $\tilde\theta_0 = 0$. We will comment more on this later. In fact, we will show that the coupling is restricted to take values in the domain $|\lambda| \leq 4$. 

For positive $\lambda$ and near $\tilde\theta = \tilde\theta_0$ the action \aadd\ is given approximately by
\eqn\aaddstarstar{\eqalign{S_k = 
{k\over 2 \pi} \int d^2z \left\{\partial r \bar\partial r + {\lambda\over 4\left(1 - {\lambda\over 4}\right)^2}\partial\tilde\varphi\bar\partial\tilde\varphi 
-\left({4 + \lambda\over 4 - \lambda}\right)\left(\bar\partial\tilde\varphi\partial\tilde\chi + \bar\partial\tilde\chi\partial\tilde\varphi\right) 
 + 2\lambda  \left({4 + \lambda\over 4 - \lambda}\right) \cdot r \cdot \bar\partial\tilde\chi\partial\tilde\chi\right\},
}
}
where $r := \tilde\theta_0 - \tilde\theta$ and it is positive. Note at $r = 0$ the metric at constant $\tilde\varphi$ has no conical singularity. However, it has a curvature singularity. Near $\tilde\theta = 0$ the action \aadd\ is given approximately by
\eqn\aaddstarstar{\eqalign{S_k = 
{k\over 2 \pi} \int d^2z \left\{\partial \tilde\theta \bar\partial \tilde\theta + {1\over 4{\tilde\theta}^2}\partial\tilde\varphi\bar\partial\tilde\varphi 
-{(4 - \lambda)\over 8{\tilde\theta}^2}\left(\bar\partial\tilde\varphi\partial\tilde\chi + \bar\partial\tilde\chi\partial\tilde\varphi\right) 
 +  {(4 - \lambda)^2\over 16 \tilde\theta^2}  \bar\partial\tilde\chi\partial\tilde\chi\right\}.
}
}
Also here in \aaddstarstar\ it is clear that there is no conical singularity at $\tilde\theta = 0$.

The shift \aacc\ does not remove the curvature singularity at $\tilde\theta = 0$. Therefore, in general the coordinate $\tilde\chi$ in \aadd\ also need not be periodic with a definite period. However, in this paper, we identify $\tilde\chi$ with $\tilde\chi + 2\pi/(1 - \lambda/4)$. Thus, $\tilde\chi$ has period $2\pi/(1 - \lambda/4)$. This choice implies that under T-duality transformation along $\tilde\chi$ the resulting metric has no conical singularity at $\tilde\theta = 0$. We will come back to this point in the next section.\foot{In the case $\tilde\chi$ in \aadd\  is periodic with period given by $2\pi$, the T-dual theory involves conical singularity which is obviously not a problem in string theory. We leave this to a future work. However, the general picture we find is independent of the period, see, \eg, Fig. 1 below.}

We finally do another T-duality along $\tilde\chi$. This leads to the single-trace $T{\bar T}$ deformed string action in global coordinates
\eqn\aaee{ S_k = {k\over 2 \pi} \int d^2z \partial Y^a \Sigma_{ab} \bar\partial Y^b,  
}
where $Y^a= (Y^0, Y^1, Y^2) = (\tilde\varphi, \tilde\psi, \tilde\theta)$ and the background fields $\Sigma_{ab}$ are
\eqn\aaff{\matrix{
\Sigma_{22} =  1,\cr
\Sigma_{11} =   \left[{1\over \sinh^2\tilde\theta} +  {\lambda\over 2}\left(-{\cosh(2\tilde\theta)\over \sinh^2\tilde\theta} +  {\lambda\over2} \left(- \cosh^2\tilde\theta  + {1\over 4}{\cosh^22\tilde\theta\over \sinh^2\tilde\theta}\right)\right)\right]^{-1} = {\sinh^2(\tilde\theta)\over 1 + \left({\lambda\over 4}\right)^2 - 2\left({\lambda\over 4}\right)\cosh(2\tilde\theta)}, &\cr
\Sigma_{01} =   -\Sigma_{10} = -\Sigma_{11}\left[-{1\over 2}{\cosh 2\tilde\theta \over \sinh^2 \tilde\theta} + {\lambda\over 2}\left( -\cosh^2\tilde\theta + {1\over 4}{\cosh^2 2\tilde\theta \over \sinh^2\tilde\theta}\right)\right] = -{1\over 2}{{\lambda\over 4} - \cosh(2\tilde\theta)\over1 + \left({\lambda\over 4}\right)^2 - 2\left({\lambda\over 4}\right)\cosh(2\tilde\theta)}, & \cr
\Sigma_{00} =  -\cosh^2\tilde\theta + {1\over 4}{\cosh^2 2\tilde\theta \over \sinh^2\tilde\theta} - \Sigma_{11}\left[-{1\over 2}{\cosh 2\tilde\theta \over \sinh^2 \tilde\theta} + {\lambda\over 2}\left( -\cosh^2\tilde\theta + {1\over 4}{\cosh^2 2\tilde\theta \over \sinh^2\tilde\theta}\right)\right]^2 = -{\cosh^2(\tilde\theta)\over 1 + \left({\lambda\over 4}\right)^2 - 2\left({\lambda\over 4}\right)\cosh(2\tilde\theta)}. &
}
}
The deformed two form $B$ field is now given by
\eqn\twoformBBBBB{ B = {1\over 2}\sum_{a, b = 0}^2B_{ab}dY^a\wedge dY^b = B_{01}d\tilde\varphi\wedge d\tilde\psi, \quad B_{01} = k\alpha' \Sigma_{01} = -{1\over 2}k\alpha'{{\lambda\over 4} - \cosh(2\tilde\theta)\over1 + \left({\lambda\over 4}\right)^2 - 2\left({\lambda\over 4}\right)\cosh(2\tilde\theta)}.
}
The three form flux $H$ is then obtained by taking exterior derivative. We get $H = dB = H_{012}d\tilde\varphi\wedge d\tilde\psi\wedge d\tilde\theta$, where $H_{012} = \partial B_{01}/ \partial\tilde\theta$. It is given by
\eqn\aaeeTHREFORMMM{ H_{012} = k\alpha'{\left[1 - \left({\lambda\over 4}\right)^2\right]\sinh(2\tilde\theta)\over \left[1 + \left({\lambda\over 4}\right)^2 - 2\left({\lambda\over 4}\right)\cosh(2\tilde\theta)\right]^2}.
}
The dilaton $\phi$ is now given by
\eqn\aaeeDILATON{\phi = {1\over 2}\ln\left|{g^2_s\over 1 + \left({\lambda\over 4}\right)^2 - 2\left({\lambda\over 4}\right)\cosh(2\tilde\theta)}\right|,
}
where $g_s$ is the string coupling. The coupling satisfies the condition $\lambda \leq 4$ (since $\tilde\theta$ is real and positive, \eg, in the case $\lambda$ is positive $0 \leq \tilde\theta \leq \tilde\theta_0 = -\ln(\lambda/4)/2$). In general, it is restricted to take values in the (fundamental) domain $|\lambda| \leq 4$.  For details see Appendix E. We will comment further on this later in section four. The coordinate $\tilde\psi$ in \aaee\ has period $2\pi (1 - \lambda/4)$. The $S^3 \times {\cal M}_4$ part of the action is not affected by the deformation since we are only considering deforming the AdS part.

The series of transformations we performed to get the $T{\bar T}$ deformed AdS background \aaff\ are summarized pictorially in Fig. 1. The first and the last transformations, \ie, the T-duality transformations that take us from plot $a)$ to $b)$ and from plot $c)$ to $d)$, are symmetries and thus do not change the respective theories physics. The second transformation, \ie, the shift that takes us from plot $b)$ to $c)$, however, changes the boundary conditions and/or periodicities of the coordinates, see \Asrat. The plots are at constant $\tilde\varphi$. It is important to note that since the specetime ends on the boundary which is located at $\tilde\theta = \tilde\theta_0$, there are no closed timelike curves in single trace $T{\bar T}$ deformation.

\ifig\loc{$a)$ is a plot of the AdS space at constant $\tilde\varphi$. The compact direction is $\tilde\psi$. The distance from the symmetry axis to the surface is $\sinh(\tilde\theta)$, see \aaaa. Along the symmetry axis, the values of $\tilde\theta$ increases from $0$ to $+\infty$. Plot $b)$ denotes the space after making a T-duality transformation. The compact direction is the dual coordinate $\tilde\chi$, see \aabb. The distance from the symmetry axis is $1/\sinh(\tilde\theta)$. Plot $c)$ denotes the space with the metric \aadd\ obtained after making the shift \aacc. For positive coupling $\lambda$, this brings the tip of the surface from infinity to a finite value of $\tilde\theta$ given by $\tilde\theta_0 = -{1\over 2}\ln \left({\lambda\over 4}\right)$. Thus, $\lambda \leq 4$. For negative coupling $\lambda$, the tip remains at $\tilde\theta = +\infty$. However, the rate at which the surface contracts to the tip is slower compared to $b)$. Plot $d)$ denotes the space \aaff\ after making the final T-duality transformation. The compact direction is $\tilde\psi$. The distance from the axis of the symmetry is $(e^{\phi}/g_s)\sinh(\tilde\theta)$, where $\phi$ is the dilaton \aaeeDILATON. In the plots $c)$ and $d)$ we used the value $\lambda = 2$.}
{\epsfxsize5.5in\epsfbox{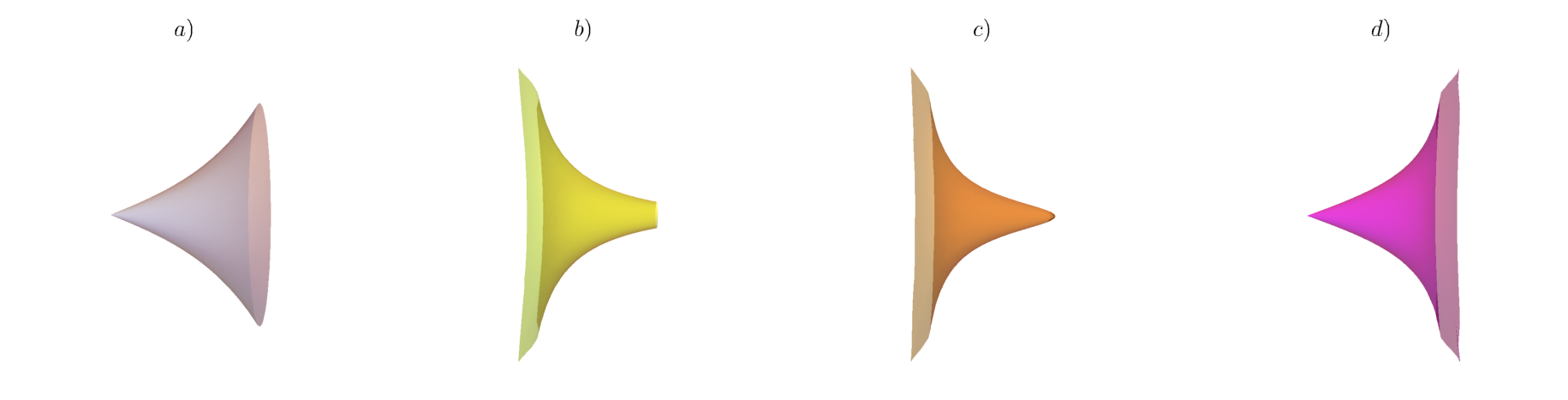}}

The action in Poincar\'e coordinates is obtained (as usual) by taking the large $\tilde\theta$ limit of \aaee. To be precise, we take $\rho := e^{\tilde\theta}$ to infinity and $\lambda$ to zero but we keep $\lambda \rho^2$ arbitrary but finite.\foot{Upon taking, in the Ricci scalar in global coordinates, this limit, \ie, we take $\rho\to \infty,\ \lambda\to 0,\ \lambda\rho^2\to {\rm arbitrary \ but \ finite}$, we obtain the correct Ricci scalar in Poincar\'e coordinates.} Thus, $\lambda \rho^2$ provides a natural coordinate to work with. Upon taking this limit, the single-trace $T{\bar T}$ deformed string action becomes
\eqn\aagg{S_k = {k\over 2 \pi}\int d^2z \left[\partial\tilde\theta\bar\partial\tilde\theta  + {1\over -\lambda + 4e^{-2\tilde\theta}} \partial (\tilde\psi + \tilde\varphi) \bar\partial (\tilde\psi - \tilde\varphi)\right],
}
which is in agreement with \refs{\Giveon, \ \IsraelM, \ \Forste}.\foot{An action identical to \aagg\ can be also obtained by performing the TsT transformation on $AdS_3$ in Poincar\'e coordinates. For this reason we sometimes refer the coordinates in \aagg\ as Poincar\'e coordinates.}  Note that the period $2\pi (1 - \lambda/4)$ of $\tilde\psi$ becomes, upon taking the limit, $2\pi$.\foot{In poincar\'e coordinates we uncompactifiy $\tilde\psi$.} The dilaton field depends only on the radial coordinate $\tilde\theta$. The exact expression is obtained making use of \aaeeDILATON\ or  (A.9). After taking the limit we get
\eqn\aaeeDILATONNN{ e^{2\phi} = {g^2_s\over 1- \left({\lambda\over 4}\right)e^{2\tilde\theta}},
}
Here in general $\tilde\theta$ runs along the real line since \aagg, \ie, the metric (with the dilaton and two-form fields), is an exact (in $\alpha'$) solution to string theory (equations of motion). We will be precise in section five. The Ricci scalar of the metric (see \aagg) is given by\foot{The Ricci scalar has mass dimension two. Here we set $k\alpha'$ to one.}
\eqn\pqaaqq{R = -32{3 + \lambda\rho^2\over (-4 + \lambda \rho^2)^2},
}
where $\rho:= e^{\tilde\theta}$ is positive. We observe that the Ricci scalar $R$ depends on $\rho$ only through the combination $\lambda\rho^2$. The components of the metric can be also made to depend on $\rho$ only through the combination $\lambda\rho^2$ by suitable redefinition of coordinates. We will use this crucial observation later in the paper. Note that the null energy condition (NEC) requires $\lambda \leq 0$ \refs{\Freedman\Girardello-\Gubser}. In a large class of theories, in general, violation of the NEC implies instability \Dubovsky. Note also that the norm of the coupling $\lambda$ can be set to one by shifting the radial coordinate.\foot{And by rescaling $\tilde\varphi,\ \tilde\psi$.} In a $T{\bar T}$ deformed CFT quantized on a circle this corresponds to changing the size of the circle. We denote the deformed part of the spacetime as ${\cal A}_3$. We next take, in the following section, the Penrose limit of the deformed string background ${\cal A}_3 \times S^3$. The manifold ${\cal M}_4$ will not play any role in our discussion and thus we simply ignore it.

\newsec{Penrose limit of $T{\bar T}$ deformed $AdS_3\times S^3$}

In this section we study the Penrose limit of the six dimensional spacetime ${\cal A}_3\times S^3$ which results in a plane wave background. Every non-pathological or physically reasonable spacetime has a Penrose limit \refs{\Penrose, \ \Berenstein}. The limit restricts the Hilbert space of the theory to a specific sector. We in particular consider the Berenstein, Maldacena and Nastase (BMN) plane wave limit \Berenstein. 

The ${\cal A}_3$ part of the metric is given by
\eqn\aahhx{\eqalign{ds_{{\cal A}_3}^2  & = k\alpha'g_{\mu\nu}dY^{\mu}dY^{\nu}, \quad g_{\mu\nu} = {\Sigma_{\mu\nu} + \Sigma_{\nu\mu}\over 2},\cr
 & = k\alpha'\left(d\tilde\theta^2 - e^{2\phi}\cosh^2\tilde\theta d\tilde\varphi^2 + e^{2\phi}\sinh^2\tilde\theta d\tilde\psi^2\right),
}
}
where the background field $\Sigma_{\mu\nu}$ is given by \aaff, and $Y^{\mu} = (Y^0, Y^1, Y^2) = (\tilde\varphi, \tilde\psi, \tilde\theta)$. The dilaton $\phi$ is given by \aaeeDILATON.\foot{Here we set $g_s = 1$ for simplicity and convenience.} The $S^3$ part of the metric is given by
\eqn\aahh{ds_{s^3}^2 = k\alpha'\left(d\theta^2 + \cos^2\theta d\varphi^2 + \sin^2\theta d\psi^2\right).
} 

The plane wave background is obtained by expanding the metric around, or by zooming in on, the geodesic of a light, or nearly so, particle that is moving along the equator of $S^3$ given by $\theta = 0$, which is the $\varphi$ direction, and positioned in ${\cal A}_3$ at $\tilde\theta = 0$. At $\tilde\theta = 0$ the metric \aahhx\ reduces to $ds_{{\cal A}_3}^2 = k\alpha' \Sigma_{00}d\tilde\varphi^2$. The null geodesic equation can be solved by $\varphi = \sqrt{-\Sigma_{00}}\cdot \tau + {\rm const.}$ and $\tilde\varphi = \tau$, where $\tau$ is an affine parameter.

We now focus on or blow up the geometry near the geodesic. To do this we introduce the coordinates,
\eqn\aaii{\tilde\theta = {r\over\sqrt{ k\alpha'}}, \quad \theta = {y\over \sqrt{k\alpha'}},
}
and take the large volume limit, \ie, $k\to \infty$. 

We first redefine $\varphi$ by rescaling it as $\varphi \to \varphi/(1 - {\lambda\over4})$.
Thus, in the BMN limit we find
\eqn\aajj{\eqalign{ds^2 = - {2k\alpha' \over (1 - {\lambda\over 4})^2}d \tilde x^-dx^+ -{1\over (1- {\lambda\over 4})^2}\left[ {(1 + {\lambda\over 4})^2\over (1 - {\lambda\over 4})^2}r^2 + y^2 \right](d{x^+})^2 + dr^2 + {1\over (1 - {\lambda\over 4})^2}\cdot r^2d\tilde\psi^2\cr
+ dy^2  + y^2d\psi^2 - k\alpha'\left[{1\over {(1 - {\lambda\over 4})^2}} + {(1 + {\lambda\over 4})^2\over (1 - {\lambda\over 4})^4}\cdot {r^2\over k\alpha'}\right]\left(d{\tilde x}^-\right)^2 - 2{(1 + {\lambda\over 4})^2 \over (1 - {\lambda\over 4})^4} r^2 d \tilde x^-dx^+ + \cdots,
}
}
where 
\eqn\aakk{ {\tilde x}^- = \tilde\varphi - \varphi, \quad x^+ = \varphi,
}
and the ellipsis consists of terms sub-leading in $k$. After rescaling ${\tilde x}^-$ and $x^+$ \foot{That is, we redefine $\tilde x^-$ by making the replacement ${\tilde x}^- \to  (1 -{\lambda\over 4} ){\tilde x^-}$. Similarly, $x^+ \to  (1 -{\lambda\over 4} )x^+$.}, the six dimensional plane wave geometry takes the form 
\eqn\aallCONIC{ds^2 = -2dx^-dx^+ -\left( \mu^2 r^2 + y^2 \right)d{x^+}^2 + dr^2  + {1\over (1 - {\lambda \over 4})^2}\cdot r^2d\tilde\psi^2+ dy^2  + y^2d\psi^2,
}
where $x^- = k\alpha' \tilde x^-$ and the parameter $\mu$ depends on the deformation coupling $\lambda$. $\mu$ is given by
\eqn\aamm{\mu^2 := {(1 + \gamma)^2\over (1 - \gamma)^2}, \quad \gamma =  {\lambda\over 4}.
}
The metric \aallCONIC\ has conical singularity at $r = 0$ unless $\tilde\psi / (1 - {\lambda\over 4})$ is periodic with a period $2\pi$, see \Sokoloff. We redefine $\tilde\psi$ by rescaling it as $\tilde\psi \to (1 - {\lambda\over 4})\tilde\psi$. This restors or fixes the periodicity of $\tilde\psi$ to $2\pi$ and in turn avoid the conical singularity, see our discussion in section two. See also Appendix A and the discussion in \Hassan. The metric now becomes
\eqn\aall{ds^2 = -2dx^-dx^+ -\left( \mu^2 r^2 + y^2 \right)d{x^+}^2 + dr^2  + r^2d\tilde\psi^2+ dy^2  + y^2d\psi^2.
}
We note that in the case the deformation coupling $\lambda$ is zero, the plane wave background is
\eqn\aallcompare{ds^2 = -2dx^-dx^+ -\left(r^2 + y^2 \right)d{x^+}^2 + dr^2  + r^2d\tilde\psi^2+ dy^2  + y^2d\psi^2.
}

The two form $B$ field has in general the form 
\eqn\bfieldafterppformg{B  =B_{-a}d{\tilde x}^-\wedge dZ^a + B_{+a}dx^+\wedge dZ^a +  {1\over 2}B_{ab}dZ^a\wedge dZ^b,
}
where $Z^a = (\tilde\varphi, \tilde\psi, \tilde\theta, \varphi, \psi, \theta)$. In general, we can use the gauge freedom to set $B_{+a} = 0$, see \eg\ \Gueven. We first use this gauge condition and then take the BMN limit. In this limit the $B$ field (modulo a closed two-form) takes the form
\eqn\aammTWOFORMBBBBB{\eqalign{B &= B_{r\tilde\psi}dr \wedge d\tilde\psi + B_{y\psi}dy \wedge d\psi,\cr
& = -2\mu x^+ rdr \wedge d\tilde\psi  - 2x^+ydy \wedge d\psi.
}
}
The three-form flux $H$ is given by $H = dB$. In general, we can consider admixture of NSNS and RR fluxes \refs{\Gueven}.

In the BMN limit the dilaton \aaeeDILATON\ is given by
\eqn\dilatonontheplane{e^{2\phi} = {g_s^2\over (1 - \gamma)^2}.
}
It depends only on the coupling.

In the next section by applying T-duality on the plane wave geometry \aall\ we obtain G\"odel (type) universes. In superstring theory, T-duality changes the chirality of fermions. It changes, for example, type IIB string theory to type IIA string theory. In this paper, we will mainly be concerned with bosonic string theory \refs{\Sadri}.

\newsec{G\"odel universes in the $T{\bar T}$ deformed string theory}
In string theory G\"odel (type) universes are T-dual of plane wave solutions \refs{\Boyda-\Harmark}. See also \refs{\Brace, \ \IsraelMM}. In this section we obtain the G\"odel (type) universe from the plane wave spacetime \aall. The spacetime \aall\ with $\mu^2 \neq 1$ is viewed as the single-trace $T{\bar T}$ deformation of the spacetime \aallcompare. 

In this section we consider the deformed spacetime with the metric \aall. We will assume that no NSNS two form $B$ field is turned on \Harmark.\foot{In the presence of the $B$ field \aammTWOFORMBBBBB\ the spacetime \aall\ is invariant under $T$ duality (after making the trivial change $\chi \to -\chi$, where $\chi$ is the T-dual coordinate), \ie, it is self T-dual.} Note also that $H^2 = 0$. The relevant metric \aall, which we write again below for convenience, is 
\eqn\aann{ds^2 = -2dx^-dx^+ -\left( \mu^2 r^2 + y^2 \right)d{x^+}^2 + dr^2  + r^2d\tilde\psi^2+ dy^2  + y^2d\psi^2.
} 
The spacetime \aann\ is homogeneous \Sadri. The Ricci scalar $R$ is independent of $\mu$ and equals zero. We first define the coordinates
\eqn\aaoo{\tilde x^1 = r\cos\tilde\psi, \quad \tilde x^2 = r\sin\tilde\psi, \quad \tilde x^3 = y\cos\psi, \quad \tilde x^4 = y\sin\psi.
}
In these coordinates the metric becomes\foot{In the variables $\tilde x^i$, the $B$ field \aammTWOFORMBBBBB\ is given by
\eqn\bfieldnowremoveafteragain{B = -2x^+\left(\mu d{\tilde x}^1\wedge d{\tilde x}^2 +  d{\tilde x}^3\wedge d{\tilde x}^4\right).
}
The three form flux is $H = dB = -2dx^+\wedge \left(\mu d{\tilde x}^1\wedge d{\tilde x}^2 +  d{\tilde x}^3\wedge d{\tilde x}^4\right)$.} 
\eqn\aapp{ds^2 = -2dx^-dx^+ -\left[ \mu^2 \left((\tilde x^1)^2 + (\tilde x^2)^2\right) + ((\tilde x^3)^2 + (\tilde x^4)^2) \right]d{x^+}^2 + (d\tilde x^1)^2+ (d\tilde x^2)^2  + (d\tilde x^3)^2 + (d\tilde x^4)^2.
}
We further make the following change of variables\foot{In general we can make $\mu$ positive by sending $x^{\pm} \to -x^{\pm}$. Thus, we will assume $\mu$ to be positive. We will come again to this point momentarily.} 
\eqn\aaqq{\tilde x^1 + i \tilde x^2 = (x^1 + i x^2)e^{-i\mu x^+}, \quad \tilde x^3 + i \tilde x^4 = (x^3 + i x^4)e^{-i x^+}.
}
In this new set of coordinates the metric takes the form \HorowitzM
\eqn\aarr{\eqalign{ds^2 & = -2dx^-dx^+ +  \sum_{i = 1}^4(dx^i)^2 -2\sum_{i, j = 1}^4J_{ij}x^idx^jdx^+,\cr 
& = -2dx^+\left(dx^- + \sum_{i,j = 1}^4 J_{ij}x^i dx^j\right) + \sum_{i = 1}^4 \left(dx^i\right)^2,\cr
& = -2dx^+\left(dx^- - \sum_{i = 1}^4 b_{i+} dx^i\right) + \sum_{i = 1}^4 \left(dx^i\right)^2,\quad b_{i+} := \sum_{j = 1}^4J_{ij}x^j,
}
}
where $J_{ij} = -J_{ji}$ and the nonzero entries of $J_{ij}$ are\foot{In the new variables, the $B$ field \bfieldnowremoveafteragain\ is now given by (modulo a closed two-form) 
\eqn\bfieldnowremoveafter{B = -J_{ij}x^idx^j\wedge dx^+,
}
where summation over repeated indices is assumed.} 
\eqn\aaqqXX{J_{12} = -J_{21} = \mu, \quad J_{34} = -J_{43} = 1.
}
For convenience, in what follows, we work with dimensionless coordinates. We redefine $x^-, x^i$ as $x^-/l_s^2, x^i/l_s$. We define  
\eqn\aarrr{x^+ = t + w, \quad 2x^- = t - w,
}
and rewrite the metric as
\eqn\aass{ds^2 = -dt^2 + dw^2 +  \sum_{i = 1}^4(dx^i)^2 -2\sum_{i, j = 1}^4J_{ij}x^idx^j(dt + dw).
}
We note that the coordinate $w$ is spacelike and an isometry of the spacetime. Thus, we can perform T-duality along $w$ \KiritsisM. This gives the G\"odel (type) universe
\eqn\aatt{ds^2 = -\left(dt + \sum_{i, j = 1}^4J_{ij}x^idx^j\right)^2 +  \sum_{i = 1}^4(dx^i)^2 + dq^2,
}
and the two-form $B$ field (modulo a closed two-form) given by
\eqn\godelbfield{B = B_{iq} dx^i\wedge dq, \quad  B_{iq} = -J_{ij}x^j,
}
where summation over repeated indices is assumed. The three-form flux is given by $H = dB = \sum_{i,j = 1}^4H_{ijq}dx^i\wedge dx^j\wedge dq$ and $H_{ijq} = J_{ij}$. The curvature scalar is not invariant under T-duality. The curvature scalar of the metric \aatt\ is $R = 2(1 + \mu^2)$. The dilaton is given by $e^{2\phi} = g_s^2/(1 - \gamma)^2$, see \dilatonontheplane.

The G\"odel type universe \aatt\ exhibits the essential properties or attributes of the four dimensional G\"odel universe obtained by Kurt G\"odel.\foot{The three dimensional subspace of G\"{o}del universe metric obtained by Kurt G\"{o}del is given by, see \Hawking
\eqn\aauusinceaskedd{ds^2 = 2\omega^{-2}\left(-dt^2 + dr^2 - \left(\sinh^4 r - \sinh^2 r\right)d\phi^2 + 2\sqrt{2}\sinh^2 r d\phi dt\right),
}
where $-\infty < t < \infty, 0 \leq r < \infty$, and $0\leq \phi \leq 2\pi$. $\phi = 0$ is identified with $\phi = 2\pi$. The Ricci scalar $R = 2\Lambda$, where $\Lambda$ is the cosmological constant. It has the value $\Lambda = -\omega^2$. It describes a rotating spacetime ($g_{\phi t} \neq 0$) filled with noninteracting dust particles of positive energy density $\rho = -R/\kappa$, where $\kappa = 8\pi G_{N}$ is the gravitational constant. The speed of light $c$ is set to $1$.} In particular, we expect the universe \aatt\ to contain light-like surface and closed timelike curves. The space \aatt\ is homogeneous, \ie, no point is preferred and any point can be reached from any other point by (a sequence of) symmetry transformations \GauntlettNM. It has the following Killing vector fields which act transitively.
\eqn\killllmegodel{
V_q = \partial_q, \quad V_t = \partial_t, \quad B_i = \partial_i - \sum_{j = 1}^4J_{ij} x^j\partial_t,
}
where $\partial_q = \partial/\partial q,\ \partial_t = \partial/\partial t$, and $\partial_i = \partial/\partial x^i$. It also has the Killing vector fields
\eqn\killllmegodelLLLL{
L_{12} = x^1\partial_2 - x^2\partial_1, \quad L_{34} = x^3\partial_4 - x^4\partial_3,
}
which generate a $U(1)\times U(1)$ symmetry, see footnote 31. The non-trivial commutation relations are
\eqn\killllmegodelCOMU{
[B_i, B_j] = 2J_{ij}V_t, \quad [L_{ij}, B_k] = -\sum_{l = 1}^4(\delta_{ik}\delta_{jl} - \delta_{jk}\delta_{il})B_l.
}
We hope to study the causal structure of the spacetime \aatt\ in detail in a future work. We will take, without loss of generality, $J_{12}$ to be positive. That is, we have 
\eqn\zzaa{
-1 \le \gamma \le 1,
} 
see \aamm\ and \aarr.

The interval \zzaa\ covers the whole positive $\mu$ line. All other values of the coupling can be mapped onto the domain \zzaa, see Appendix E. In this domain $\gamma$ is connected to zero continuously. As we will discuss momentarily, the restriction \zzaa\ is related to the curvature singularity of the underlying string theory \aagg, \eg\ see the Ricci scalar \pqaaqq. It can also be understood by examining the sign of the $\bar\partial\tilde\chi\partial\tilde\chi$ coefficient in \aadd. 

We note that the section of the universe in which $x^1 = {\rm const.}$ and $x^2 = {\rm const.}$ is independent of the coupling, and thus, there are string states that do not deform. These are string states that live on the sphere. For simplicity and convenience we consider the section of the universe where $x^3 = {\rm const.}$ and $x^4 = {\rm const.}$. It has the topology of a cylinder times a circle which we uncompactify. The line subspace is parametrized by $q$. In what follows we will consider an observer positioned at $x^3 = {\rm const.}$ and $x^4 = {\rm const.}$.

Therefore, for constant $x^3$ and $x^4$, the G\"{o}del (type) universe metric \aatt\ is four dimensional and it is the direct sum of the metric on the manifold with coordinates $(t, x^1, x^2)$ which is given by
\eqn\aauu{ds^2 = -\left[dt + \mu\left(x^1dx^2 - x^2dx^1\right)\right]^2 +  \sum_{i = 1}^2(dx^i)^2,
}
and the metric on the line which is given by
\eqn\aauusinceasked{ds^2 = dq^2.
}
Therefore, we set, without loss of generality, $q = 0$, \ie, we will simply ignore the line subspace in our following discussion as it is also done in \Hawking\ for obvious reason. In order to describe the essential characteristic properties of the universe under single-trace $T{\bar T}$ deformation considering this section is sufficient since the spacetime \aatt\ is homogeneous \refs{\GauntlettNM,\ \Kowalski}, and other choices do not alter the conclusions.\foot{This section corresponds equivalently to setting $y$ to zero, see equations \aaii, \aaoo. This can alternatively be viewed as setting in \aaqqXX\ $J_{34} = -J_{43} = 0$.} In cylindrical coordinates 
\eqn\aavv{x^1 = r\sin\xi, \quad x^2 = r\cos\xi,
}
the metric \aauu\ becomes\foot{In general we can put the relevant part of the metric \aatt\ into the form
\eqn\surfacelight{\matrix{ds^2  =  -\left(dt - \mu \left(r^2_1 d\varphi_1 + r_2^2 d\varphi_2\right)\right)^2 + dr_1^2 + r_1^2d\varphi_1^2 + \mu\left(dr_2^2 + r_2^2d\varphi_2^2\right),\cr
 = -\left(dt - r^2\left(\mu\cos^2\theta d\varphi_1 + \sin^2\theta d\varphi_2\right)\right)^2 + dr^2 + r^2\left(d\theta^2 + \cos^2\theta d\varphi_1^2 + \sin^2\theta d\varphi_2^2\right),
}}
where $0 \leq r_1,r_2 < \infty$, and $0\leq \varphi_1,\varphi_2 \leq 2\pi$. $ 0\leq r < \infty$ and $0\leq \theta\leq \pi/2$. $\theta = 0$ is a great circle on the three unit sphere, and it is the case $x^3 = {\rm const.}$ and $x^4 = {\rm const.}$.

In terms of the left and right invariant Maurer-Cartan one-forms of the three sphere we can write the metric as
\eqn\forminvariantuone{ds^2 = -\left[dt - r^2\left(\mu_- \omega_3 + \mu_+\bar\omega_3\right)\right]^2 + dr^2 + r^2d\Omega_3^2,
}
where $d\Omega_3^2 = {1\over 4}\sum_{i = 1}^3\omega_i^2 = {1\over 4}\sum_{i = 1}^3\bar\omega_i^2$ is the metric on the unit three sphere. $\mu_\pm = (\mu \pm 1)/4$. $\omega_i$ are the left invariant one-forms and $\bar\omega_i$ are the right invariant one-forms, see Appendix D. The form \forminvariantuone\ clearly manifests the $U(1)\times U(1)$ symmetry. We note that in the case $\mu_- = 0$ the symmetry enhances to $SU(2)\times U(1)$.
}
\eqn\aaww{ds^2 = -\left(dt - \mu r^2 d\xi\right)^2 +  dr^2 + r^2d\xi^2.
}

It is known that every homogeneous spacetime admits a geodesic which is an orbit of a one-parameter group of isometries \Kowalski. Therefore, consider the curve $\gamma(s) = \{(t(s), r(s), \xi(s))\}$ with $t = {\rm const.}$ and $r = {\rm const.}$. Here $s$ is a point in $\IR$ (or an interval in $\IR$). The norm of the Killing vector $L_{12} := v = ((\partial/\partial\xi)^\mu) = (0, 0, 1)$ which is tangent to this curve is
\eqn\aaxx{v^2 := v_\mu v^\mu = g_{\xi\xi} = r^2(1 - \mu^2 r^2).
}
For a spacelike tangent vector $v^2 > 0$. Since $v^2 \le 0$ for $r \ge r_c$, where $r_c$ is the critical radius given by
\eqn\aayy{r_c := 1/\mu,
}
there exists closed timelike and null curves. However, we stress that there are no closed timelike and null geodesics. The critical radius is the outermost radial point a geodesic which started at the origin can reach. The G\"{o}del spacetime is homogeneous.\foot{Note that, on the contrary to the usual assumption in cosmological models, the spacetime is not isotropic.}  Thus, for every point of the spacetime a closed timelike curve can be found that passes through it. However, every closed timelike curve must intersect the outer region $r > r_c$.

Light rays emitted from the origin follow a spiral trajectory \Hawking. In a finite affine parameter they reach the surface at $r = r_c$. The surface at $r_c$ is a trivial bundle over a null circle and it is referred to as velocity-of-light surface. Then, the light rays begin to reconverge and, in the same affine parameter, they reach the origin. The behavior of timelike geodesics is similar. They reach a maximum value of $r$ less than $r_c$ and then reconverge at the origin with same affine parameter. Light emitted from beyond the critical radius $r > r_c$ can never reach the origin.

The holographic screen is obtained by maximizing the area of constant $t$ and $r$ \Bousso. The area is one dimensional and it is given by
\eqn\aazz{A(r) = 2\pi \cdot r(1 - \mu^2 r^2)^{1\over 2}.
}
This gives the location of the holographic screen
\eqn\bbaa{r_{\cal H} = {1\over 2^{1\over 2}\mu} = {1\over 2^{1\over 2}}r_c,
}
which is less than the critical radius. Therefore, the closed timelike curves are either hidden from the observer by the screen or they are cut into causal parts. The area \aazz\ at the holographic screen radius $r_{\cal H}$ is given by
\eqn\zzbbx{A(r_{\cal H}) = {\pi\over \mu}.
}
This gives a bound on the entropy $S$ \refs{\Bousso,\ \Bekenstein,\ \Wald}. It is given by
\eqn\zzbb{S = {A(r_{\cal H})\over 4} = {\pi\over 4\mu}.
}
 We note that $S$ is positive and at $\gamma = 1$ it is zero. The angular speed $\Omega_{{\cal G}}$ with which G\"odel (type) universe rotates is given by \refs{\Godel, \ \Buser},
\eqn\qzzb{\Omega_{{\cal G}} := {1\over r_{\cal{H}}}.
}
Thus, $r_{\cal H}$ determines the intrinsic rotation (or vorticity) of the universe and also defines the holographic screen. In string units, we have $l_s = 1/2^{1\over 2}$, thus, we can write $r_{\cal H} = (1/\mu) l_s.$\foot{For convenience one may find it is useful to think of $r_{\cal H}$ as the radial location of the AdS boundary. This will become more evident momentarily.} 

Consider the case in which the coupling $\gamma$ is positive \zzaa. That is, $0\leq \gamma \leq 1$. Thus, in this case as $\gamma$ becomes larger and larger, $\mu$ goes to infinity, and hence $r_{\cal H}$ goes to zero. The holographic screen moves into the bulk. The other case is when the coupling $\gamma$ is negative. In this case $-1 \leq \gamma \leq 0$. We note that as the norm of $\gamma$ gets larger and larger, $\mu$ goes to zero, and hence $r_{\cal H}$ goes to infinity. Thus, the holographic screen moves away from the bulk. That is, it pushes the closed timelike curves further away to infinity. Therefore, in view of these facts and the uniqueness/factorization property (see \eg, \refs{\Aharony, \ \AsratT}), in holography, it is natural to view the single-trace $T{\bar T}$ deformation as moving the boundary either into the bulk or out to infinity. In what follows, we explain these observations and draw some conclusions on single-trace $T{\bar T}$ deformation. For simplicity we will mainly use the Poincar\'{e} coordinates \aagg. 

\newsec{Moving holographic boundaries with $T{\bar T}$}

We noted earlier in section two that, depending on the sign of the coupling, we have two inequivalent classes of string theory backgrounds, see \aagg. However, once we decide which sign to consider, the value of the coupling can be fixed to a preferred value, \ie, $\gamma = \pm 1$, by shifting the radial coordinate $\tilde\theta$. In this section we use the observations made in the previous section to better understand the (single trace) $T{\bar T}$ deformation.

First, let us consider the case in which $\gamma$ approaches or equals zero. The boundary (or spacetime) theory is conformal. The holographic boundary is located at large $\rho$.

We recall that in general $\rho$ enters in the Ricci scalar (see \pqaaqq) through the combination $\gamma \rho^2$. Thus, in general the natural variable or coordinate to work with or better suited is not $\rho$, but $\gamma\rho^2$. Recall that we obtained the metric (see \aagg)
\eqn\metricAdSg{ds^2 = k\alpha'\left[{d\rho^2\over \rho^2} + {1\over 4} \cdot {\rho^2\over 1 - \gamma \rho^2}\left(-d\tilde\varphi^2 + d\tilde\psi^2\right)\right],
}
by taking $\gamma \to 0$, $\rho \to \infty$ and keeping $\gamma \rho^2$ arbitrary but finite. Since the Ricci scalar of AdS is constant (and negative), at the boundary we in particular also have $|\gamma|\rho^2 := b_{0} = {\rm finite\ const.}$, see \pqaaqq.\foot{Putting back the dimensions, this is $|{\hat\gamma}|{\hat\rho}^2 = b_{0}$ where $\hat\rho = \rho/l_s$ and $\hat\gamma = \gamma l_s^2$.}

In general the constant $b_0$ can take any positive value. We note that for small $b_0$ the metric \metricAdSg\ can be put (up to an overall positive constant) into the form
\eqn\metricAdS{ds^2 = k\alpha'\left[{dr^2\over r^2} + r(-dt^2 + dx^2)\right], \quad 0 \leq r \leq b_0 \ll 1,
}
where $r = |\gamma| \rho^2,\ t = {\tilde \varphi}/\sqrt{|\gamma|},\ x = {\tilde \psi}/\sqrt{|\gamma|}$.\foot{To get $AdS$ spacetime one can simply set $\gamma \rho^2 = b_0$ and rescale the coordinates $\tilde\varphi$ and $\tilde\psi$ by $\sqrt{1 - b_0}$. This leads to the usual Poincar\'{e} metric. $\rho$ is the usual scale factor and $b_0$ is the location of the conformal boundary.} The Ricci scalar $R = -3/2$. Note $R$ does not depend on $r$ and thus $b_0$. However, since \metricAdS\ is an $\alpha'$ exact solution to string theory we can extend the range or the value of $b_0$ to one. Therefore, we set $b_0 = 1$.\foot{In general $b_0$ can take any positive finite value, specifically it is given by the pole of the Ricci scalar $R$ (see \pqaaqq). In general $\gamma = \alpha\cdot\lambda$ for some positive real number $\alpha$, see the discussion on the positive coupling case below.} Thus, at the boundary $\rho = \lim_{|\gamma|\to 0}1/\sqrt{|\gamma|}$. It is (useful and convenient to think of $b_0$ as) the position of the (conformal) boundary in string units. The coordinate $r$ has a natural interpretation as ``conformal" coordinate. This is complementary to our comment in footnote $7$.

In the case in which the coupling $\gamma$ is negative, we note that the deformed string metric \metricAdSg\ is well-defined and has a non-singular Ricci scalar \pqaaqq. We choose, without loss of generality, $\gamma = -1$; this is our preferred value. We can always achieve this by redefining the coordinates.

We note that this choice of value for $\gamma$ corresponds, in our discussion in the earlier section, to placing G\"odel's holographic screen at infinity, see \bbaa. Thus, at $\gamma = -1$ there are no closed timelike curves in the subspacetime \aauu. In general it corresponds to a slowly rotating G\"odel (type) universe \qzzb. Thus, in view of this, the timelike boundary of AdS is washed away to infinity. The boundary is at $-\gamma\rho^2 : = b_{-1} = \infty$ in string units. 

For positive coupling, on the other hand, the Ricci scalar (see \pqaaqq) is always singular. Thus, for positive values of the coupling $\gamma$, the boundary\foot{ At $\gamma = 0$ the boundary is at $\gamma\rho^2 = b_0 = 1$ in string units.}  turns into a singularity. Therefore, at the boundary or singularity, we have $\gamma \rho^2 = 1$. Thus, as we increases $\gamma$, $\rho$ decreases, see plot $c)$ in Fig. 1. Note we are not assuming a specific value for the coupling. It can take any positive value.\foot{In global coordinates $\gamma < 1$ since the radial coordinate is positive.} We expect this also to be the case in the usual $T{\bar T}(x)$ deformation as well, since, as we explained in the introduction, in the untwisted long string sector the single trace and ordinary $T{\bar T}(x)$ deformations are the same.


In general a positive coupling $\gamma$, as we noted in the discussion in the earlier section, corresponds to a relatively fast rotating G\"odel (type) universe \qzzb. As we increase $\gamma$ to one, the holographic screen moves towards the origin. Thus, at $\gamma = 1$ it terminates or ceases to exist and therefore, the G\"odel (type) universe becomes pathological \Hawking.

Had we taken the Penrose limit around a different $\tilde\theta$ value or null geodesic, the interval \zzaa\ which we write here again for convenience 
\eqn\pzzaa{-4 \leq \lambda \leq 4,
}
and the value of $\lambda$ at which the holographic screen ceases to exist, \ie, $\lambda = 4$, would have been different. That is to say, the coupling $\lambda$ can take any value.\foot{In global coordinates $\gamma$ is less than 1 since $\tilde\theta$ is positive. In Poincar\'e coordinates, however, $\tilde\theta$ can take negative values, see \eg\ \aagg, thus $\gamma$ is not restricted to a certain region on the real line.} Therefore, in general, positive values of the coupling $\lambda$ do not lead to a sensible theory. However, we still would have the same interval \zzaa\ for $\gamma$, except that we would now have to normalize $\lambda$ differently. The value of $\gamma$ at which the holographic screen terminates or dissolves would still also be the same, \ie, $\gamma = 1$. Thus, we still would have the same conclusions.

To see why the (fundamental) interval \zzaa\ is the correct domain for the coupling consider the following transformations and/or redefinitions. 
\eqn\changesymm{\gamma \to {1\over \gamma}, \quad \tilde\varphi \to {1 \over \gamma}\tilde\varphi, \quad \tilde\psi \to -{1\over \gamma}\tilde\psi.
}
The transformations \changesymm\ together with the gauge freedom or ambiguity in choosing the Kalb-Ramond $B$ field leave the action \aaee\ invariant. Therefore, in general the coupling $\gamma$ takes its value in the domain \zzaa, see Appendix E.

It is interesting to note that the presence of the curvature singularity (see \aagg) is captured in the T-dual of the BMN limit by the dissolution of the holographic screen at $\gamma = 1$. It is also interesting to note that since $\gamma$ can take any value between zero and one (see \zzaa), and also since they give equivalent string backgrounds (see \aagg), the discussion above implies, moving the AdS boundary into the bulk creates a curvature singularity. In this case the timelike boundary ceases to exist, \ie, it dissolves. Therefore, the metric \aagg\ loses its physical meaning in the same way the G\"odel (type) universe does, \ie, the holographic screen closes in on the observer and the universe becomes empty (and pathological).

In general, to make the theory sensible and avoid the curvature singularity we need to introduce a UV cutoff. We require $\rho \le \rho_0 < e^{\tilde\theta_0}$ where $\tilde\theta_0 := -{1\over2}\ln\gamma$. We also need to provide suitable boundary conditions for the bulk fields at $\rho_0$. The radial position of the boundary is $\gamma\rho_0^2 := b_{1}$ in string units. Thus, we have $b_{1} < b_0$. Therefore, this way single-trace $T{\bar T}$ deformation is said to move the boundary into the bulk.

These observations are in agreement with one's expectations. In the case $\gamma \leq 0$ we expect the deformed spacetime theory to be free of singularities and/or ambiguities. In particular, since the NEC is satisfied, we expect the energies to be real.\foot{One exception to this, \eg, is the Casimir vacuum \Casimir.} Indeed, for negative coupling, the spectrum of the deformed spacetime theory dual to the long string sector (quantized on a circle) is real, see \Asrat\ and references therein. However, the boundary (radial) position $b_0$ is now washed away to infinity $b_{-1}$, \ie, $b_0 < b_{-1}  = \infty$. In the case $\gamma > 0$ we expect a pathology/feature analogous or dual to the curvature singularity. In general, for positive coupling, the spectrum is complex, see \Asrat\ and references therein. Thus, the creation of the curvature singularity is accompanied in the deformed theory by the energies of some states becoming imaginary. In this case the timelike boundary ceases to exist, \ie, it dissolves. To make the theory well-defined, in general, we may need to impose a UV cutoff, \ie, $b_1 < b_0$, and a set of suitable boundary conditions at $b_1$. However, we stress that it is not clear whether such a set exists.

\newsec{Discussion}

In this paper we studied G\"odel (type) universes \aaww\ in single-trace $T{\bar T}$ deformation to gain insights into (single-trace) $T{\bar T}$ deformation and holography. The single-trace $T{\bar T}$ deformed string background \aaee\ is obtained by performing TsT transformation on $AdS_3$. In global coordinates, to the best of my knowledge, the single-trace $T{\bar T}$ deformed string background \aaee, which we again write here for convenience, has not been reported elsewhere in the literature before. The deformed string action in global coordinates is
\eqn\summaryaction{S_k = {k\over 2 \pi} \int d^2z \partial Y^a \Sigma_{ab} \bar\partial Y^b, \quad k\alpha'\Sigma_{ab} = g_{ab} + B_{ab}, 
}
where $Y^a= (Y^0, Y^1, Y^2) = (\tilde\varphi, \tilde\psi, \tilde\theta)$, $g_{ab}$ is the metric and $B_{ab}$ is the NSNS two form field. The background fields $\Sigma_{ab}$, the dilaton $\phi$ and the NSNS three form flux $H$ are given by
 \eqn\summarymetric{\eqalign{
 \Sigma_{22} & =  1, \cr
 \Sigma_{11} & =  (e^{2\phi}/g_s^2)\sinh^2\tilde\theta,\cr
 \Sigma_{00} & =  -(e^{2\phi}/g_s^2)\cosh^2\tilde\theta,\cr
 \Sigma_{01} & =  -\Sigma_{10} = -{1\over 2}(e^{2\phi}/g_s^2)\left({\lambda\over 4} - \cosh(2\tilde\theta)\right),\cr
 B & =  B_{01}d\tilde\varphi\wedge d\tilde\psi, \quad B_{01} = k\alpha' \Sigma_{01},\cr
 H & =  dB = H_{012}d\tilde\varphi\wedge d\tilde\psi\wedge d\tilde\theta, \quad H_{012} = \partial B_{01}/\partial\tilde\theta,\cr
 e^{2\phi} & =  {g_s^2\over 1 + \left({\lambda\over 4}\right)^2 - 2\left({\lambda\over 4}\right)\cosh(2\tilde\theta)}.
}
}
The radial location of the curvature singularity is given by the zero of $e^{-2\phi}$. At $\lambda = 4$, the singularity is at $\tilde\theta = 0$. Therefore, $\lambda \leq 4$ since $\tilde\theta$ is real and positive, \eg, in the case $\lambda$ is positive $0 \leq \tilde\theta \leq \tilde\theta_0 = -\ln(\lambda/4)/2$. The coupling in general is restricted to take values in the fundamental domain $|\lambda| \leq 4$, see appendix E for details. 

Note in the proposal \McGough, which is a proposal in the usual (or double trace) $T{\bar T}$ deformation, the deformed metric is that of $AdS_3$ with a boundary condition at finite radial distance. Thus, clearly different from \summarymetric. Also note that there is no reason for the results to be similar since the deformations are in general different.

To examine the situation in which $\lambda$ is very close to $4$ or equivalently $\tilde\theta$ is very small, we first write $\gamma = \lambda/4 =1 - \epsilon^2$ and $\tilde\theta^2 = \epsilon^4(1 - \alpha^4)/4$, where $\epsilon \ll 1$. We assume, without loss of generality, $\alpha$ is, at least, of the order of $\epsilon^{1/4}$. This gives the metric $ds^2/(k\alpha') = {1\over 16 \epsilon^4}\left(\epsilon^4d\alpha^4\right)^2 - {1\over \epsilon^{4}\alpha^{4}}d\tilde\varphi^2 + {\epsilon^4\over 4}\cdot{1\over \epsilon^{4} \alpha^{4}}d\tilde\psi^2$. The $B$ field is $B_{01} = k\alpha'/(2\epsilon^2\alpha^4)$. The dilaton is given by $e^{2\phi} = g_s^2/(\epsilon^4\alpha^4)$. The background fields can be put into the form
\eqn\metricbdil{\eqalign{ds^2 & = k\alpha'\left(dr^2 - {1\over  r}dt^2 + {1\over  r}d\eta^2\right),\cr
& = k\alpha'\left(dr^2 - {dx^+dx^-\over r}\right),\cr
B & = {k\alpha'\over r} \cdot dt\wedge d\eta,\cr 
& = {k\alpha'\over r} \cdot {1\over 2} \cdot dx^-\wedge dx^+,\cr
\phi & = -{1\over 2}\ln r + \kappa_\epsilon, \quad \kappa_\epsilon = {1\over 2}\ln \left(g_s^2\over 4\epsilon^2\right),
}
}
where $r = \epsilon^2\alpha^4/4 \ll 1$, $t = \left(1/2\epsilon\right)\cdot\tilde\varphi$, $\eta = \left(\epsilon/4\right)\cdot\tilde\psi$ and $x^\pm = t \pm \eta$.\foot{
In (super)gravity the metric, flux and dilaton \metricbdil\ are governed by the action \CallanCGS
\eqn\gractionstringframe{S = {1\over 2\kappa^2}\int d^3x\sqrt{-g}e^{-2\Phi}\left[R + 4\left(\nabla \Phi\right)^2 - {1\over 12}H^2\right],
}
where $x^a = (x^0, x^1, x^2) = (t, \eta, r)$ and
\eqn\gractionstringframeparam{e^{\Phi} = {e^{\phi}\over g_s}, \quad {1\over 2\kappa^2} = {1\over 2\kappa_0^2 g_s^2} = {1\over 16\pi G_N},
}
where $G_N$ is the gravitational coupling constant in three spacetime dimensions.

The background fields \metricbdil\ solve the equations of motion that follow from the action \gractionstringframe
\eqn\eqmstringframe{\eqalign{
R_{\mu\nu} + 2\nabla_\mu\nabla_\nu \Phi -{1\over 4} H_{\mu\rho\sigma}H_{\nu}^{\rho\sigma}& = 0,\cr
\nabla^\rho\left(e^{-2\Phi}H_{\mu\nu\rho}\right) & = 0 + \varepsilon^{\alpha\beta}g_{\mu\alpha}g_{\nu\beta}{\cal O}(\epsilon^2),\cr
R + 4\nabla^2\Phi - 4\left(\nabla \Phi\right)^2 - {1\over 12}H^2& = 0,
}
}where $\varepsilon^{01} = -\varepsilon^{10} = -1$ and $\varepsilon^{21} = - \varepsilon^{12} = 0$ is the totally antisymmetric tensor.
}
Note that near $r$ equals zero $\eta$ parameterizes non-contractable loops, and thus the metric cannot have a conical singularity at $r = 0$. Indeed, the Ricci scalar is $R = -7/(2r^2) = -56/(\epsilon^4\alpha^8)$. Note for a given $\epsilon$, $\tilde\theta$ increases to $ \epsilon^2/2 \ll 1$ as we decrease $r$ to zero. The geometry the metric describes looks like a funnel, see  Fig. 2. Thus, the strings get stretched more and more as $r$ gets closer and closer to zero.

\ifig\loc{The plot depicts the space the metric $ds^2/{k\alpha'}$ \metricbdil\ describes at constant $t$. The compact direction is $\eta$. The distance from the symmetry axis to the surface is $1/r$. Along the symmetry axis $r$ increases.}
{\epsfxsize2.5in\epsfbox{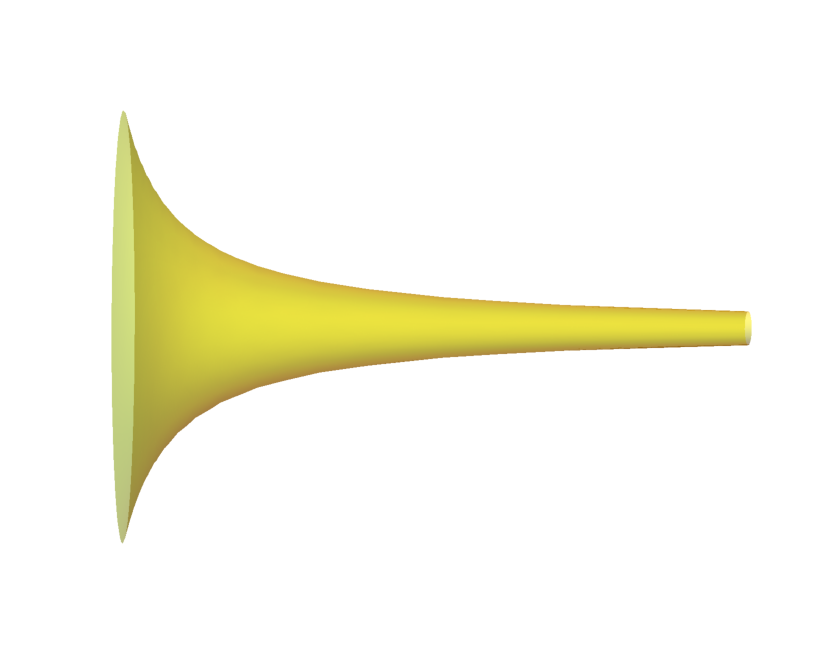}}

Note that if we naively take $\lambda = 4$ in \aaff, \aaeeTHREFORMMM\ and \aaeeDILATON\ the background fields reduce to 
\eqn\eqmstringframesol{\eqalign{
ds^2 = k\alpha'\left(d\tilde\theta^2 + {1\over 4}\coth^2\tilde\theta d\tilde\varphi^2 - {1\over 4}d\tilde\psi^2\right),\cr
 g_s^2e^{-2\phi} = 4\cdot \sinh^2\tilde\theta.
}
}
The three form NSNS flux $H$ is zero. 

Comments:
\item{1.} The metric \eqmstringframesol\ would have described the (exterior) spacetime beyond the curvature singularity had it existed or been physical, see, \eg, how the signs of the coefficients of $d\tilde\varphi^2$ and $d\tilde\psi^2$ in \aagg\ changes as we increase $\gamma\rho^2$ beyond one. To describe the singularity or ``interior" spacetime, therefore, we need to apply Wick rotations.
\item{2.} The bulk spacetime ends on the boundary. Thus, there are no closed timelike curves in single trace $T{\bar T}$ deformation.

We first apply a Wick rotation on $\tilde\psi$ to analytical continue to a Euclidean signature. There is no $B$ field. Thus, we simply set $\tilde\psi = -i\xi$. The result of the rotation and the trivial redefinition $\tilde\theta \to \tilde\theta/2$ is
\eqn\eqmstringframesolwick{\eqalign{
ds^2 = {k\alpha'\over 4}\left[d\tilde\theta^2 + \coth^2(\tilde\theta/2) d\tilde\varphi^2 + d\xi^2\right],\cr
g_s^2e^{-2\phi} =  4\cdot \sinh^2(\tilde\theta/2).
}
}
The Ricci invariant is $R = -4 \cdot {\rm csch}^2(\tilde\theta/2)$. We could compactify $\tilde\varphi$ on a unit circle such that $0 \leq \tilde\theta < \infty$, and $0\leq \tilde\varphi,\ \xi \leq 2\pi$. The metric has the topology of $D^2\times S^1$, where $D^2$ is a two-disc. At $\tilde\theta = 0$ it has a curvature singularity. Note that $\tilde\varphi$ parameterizes a non-contractable loop, and thus it need not be periodic with definite period to avoid a conical singularity at $\tilde\theta = 0$. The geometry which the metric describes looks like a funnel, see Fig. 2. It is wider at $\tilde\theta = 0$. It asymptotes for large $\tilde\theta$ to a cylinder. The T-duality transformation on $\tilde\varphi$ gives 
\eqn\cigarEU{\eqalign{
ds^2 = {k\alpha'\over 4}\left[d\tilde\theta^2 + \tanh^2(\tilde\theta/2) d\iota^2 + d\xi^2\right],\cr
g_s^2e^{-2\phi} =  4\cdot \cosh^2(\tilde\theta/2),
}
}
where $0 \leq \tilde\theta < \infty$, and $0 \leq \iota/2,\ \xi \leq 2\pi$. The Ricci invariant is $R = 4 \cdot {\rm sech}^2(\tilde\theta/2)$. We note in particular that the T-dual solution has no singularity, \ie, the curvature and dilaton are regular, however it has a horizon \GiveonMMMMMM. The manifold that the metric describes at constant $\xi$ looks like a semi-cigar \refs{\WittenMMM,\ \Mandal}, see Fig. 3. The tip of the cigar is at $\tilde\theta = 0$. Far away from the tip the manifold asymptotes to a cylinder.
\ifig\loc{The plot depicts the space the metric $ds^2/{k\alpha'}$ \cigarEU\ describes at constant $\xi$. The compact direction is $\iota$. The distance from the symmetry axis to the surface is $\tanh(\tilde\theta/2)$. Along the symmetry axis $\tilde\theta$ increases.}
{\epsfxsize2.5in\epsfbox{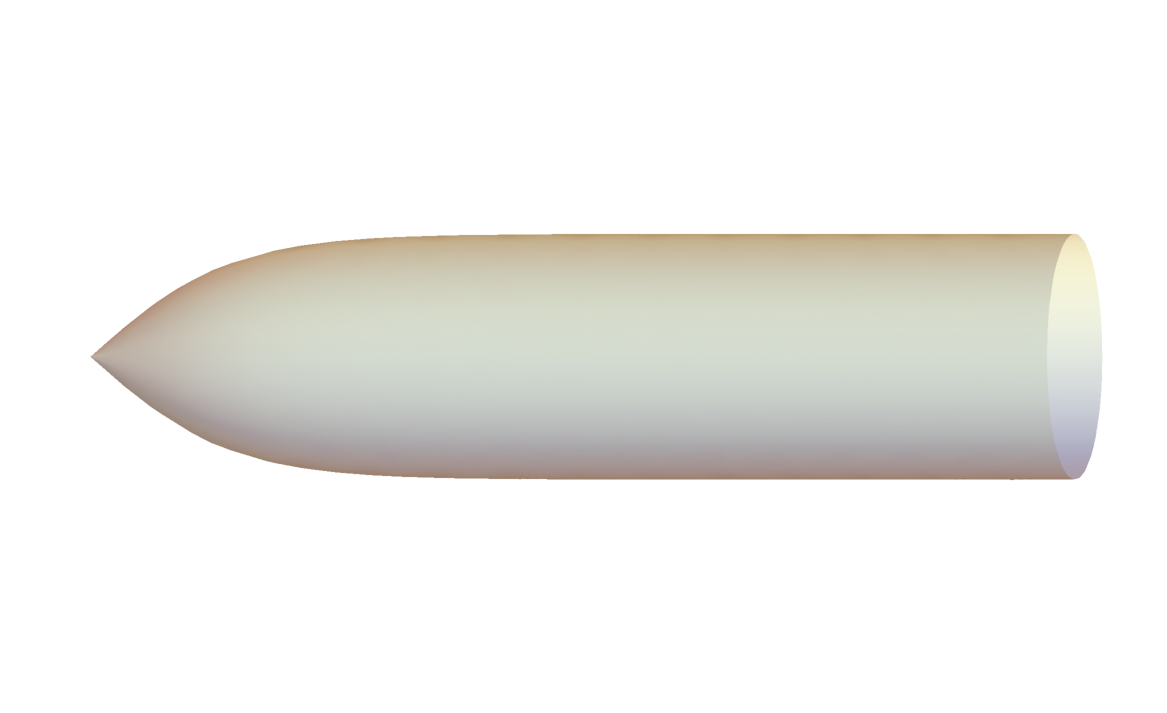}}

We note that a further Wick rotation on $\tilde\varphi$ in \eqmstringframesolwick\ gives
\eqn\eqmstringframesolwickwick{\eqalign{
ds^2 = {k\alpha'\over 4}\left[d\tilde\theta^2 - \coth^2(\tilde\theta/2) d\tau^2 + d\xi^2\right],\cr
g_s^2e^{-2\phi} =  4\cdot \sinh^2(\tilde\theta/2).
}
}
This is the same result obtained in \refs{\DijkgraafMM,\ \GinspargM,\ \SfetsosM} in the semi-classical large $k$ limit. This describes the singularity. Another way to get the result \eqmstringframesolwickwick\ from \eqmstringframesol\ is by performing the following three steps.
\item{1)} We first make a T-duality transformation along the $\tilde\varphi$ coordinate. This introduces the T-dual coordinate $\iota$. The solution we get has no curvature and/or dilaton singularity. However, it has a regular horizon. The result is
\eqn\steponetwt{\eqalign{
ds^2 = k\alpha'\left(d\tilde\theta^2 + 4\cdot \tanh^2\tilde\theta d\iota^2 -{1\over 4} \cdot d\tilde\psi^2\right),\cr
g_s^2e^{-2\phi} =  \cosh^2\tilde\theta,
}
}
where $\iota$ is periodic with period $\pi$.
\item{2)} We next apply Wick rotations on the T-dual coordinate $\iota$ and $\tilde\psi$. This is allowed since the metric and dilaton are not singular (and also there is no $B$ field). We simply set $\iota = -i\upsilon$ and $\tilde\psi = -i\xi$. This changes the metric signature. We find
\eqn\steptwotwt{\eqalign{
ds^2 = k\alpha'\left(d\tilde\theta^2 - 4 \cdot \tanh^2\tilde\theta d\upsilon^2 + {1\over 4} \cdot d\xi^2\right),\cr
g_s^2e^{-2\phi} =  \cosh^2\tilde\theta.
}
}
\item{3)} We finally apply inverse T-duality transformation along the Wick rotated T-dual coordinate $\upsilon$, see \HullMMMM. This introduces the dual coordinate $\tau$. This gives 
\eqn\stepthreetwt{\eqalign{
ds^2 = k\alpha'\left(d\tilde\theta^2 - {1\over 4} \cdot \coth^2\tilde\theta d\tau^2 + {1\over 4} \cdot d\xi^2\right),\cr
g_s^2e^{-2\phi} =  4\cdot \sinh^2\tilde\theta.
}
}
The curvature is singular at $\tilde\theta = 0$ but not of conical type and thus $\tau$ need not have a definite period. This is the result \eqmstringframesolwickwick, whereupon $\tau$ is uncompactified.

To describe, in the sense of \Codina, the interior of the singularity, we analytically continue $\tilde\theta$. This gives
\eqn\eqmstringframesolwickwicksing{\eqalign{
ds^2 = {k\alpha'\over 4}\left[-d\chi^2 + \cot^2(\chi/2) d\tau^2 + d\xi^2\right],\cr
g_s^2e^{-2\phi} =  4\cdot \sin^2(\chi/2),
}
}
where $-\pi \le \chi \le 0$. Note $\chi$ is timelike and thus \eqmstringframesolwickwicksing\ represents a cosmology. The Ricci invariant is $R = 4\csc^2(\chi/2)$. At $\chi = 0$, the dilaton and curvature diverge. Thus, the solutions \eqmstringframesolwickwick\ and \eqmstringframesolwickwicksing\ matches at $\chi = 0 = \tilde\theta$. $\chi = -\pi$ corresponds to the horizon. Therefore, the black hole solution is naked. The observation is in agreement with one's expectation and in general positive $\lambda$ leads to unphysical phenomena. Note $\chi$ and $\tilde\theta$ are different coordinates and together they do not correspond a single solution \Codina. We will discuss in detail \eqmstringframesol\ and its interior solution \eqmstringframesolwickwicksing\ in a separate paper.

In string theory, G\"odel (type) universes are (usually) obtained first by taking the Penrose limit and then by performing T-duality. The Penrose limit of the spacetime ${\cal A}_3 \times S^3$ with the metric given by the direct sum of \summarymetric\ and the metric on the three sphere \stringonsphere\ is given by \aall, which we again write below for convenience
\eqn\aallagaindiscussion{ds^2 = -2dx^-dx^+ -\left( \mu^2 r^2 + y^2 \right)d{x^+}^2 + dr^2  + r^2d\tilde\psi^2+ dy^2  + y^2d\psi^2,
}
where $\mu$ is given by \aamm. It is interesting to note that in the case the coupling $\gamma = -1$, \ie\ $\mu = 0$, the spacetime \aallagaindiscussion\ reduces to
\eqn\aallagainspecial{\eqalign{ds^2 & = -2dx^-dx^+ - y^2 d{x^+}^2 + dy^2  + y^2d\psi^2 + dr^2 + r^2d\tilde\psi^2,\cr
& = -2dx^+\left(dx^- + y^2 d\hat\psi\right) + dy^2 + y^2d\hat\psi^2 + dr^2 +  r^2d\tilde\psi^2,\cr
& = -2dx^+\left(dx^- - b_{\hat\psi +}d\hat\psi\right) + dy^2 + y^2d\hat\psi^2 + dr^2 +  r^2d\tilde\psi^2, \quad b_{\hat\psi +} := -y^2.
}
} 
Here we have made the replacement $\psi = \hat\psi + x^+$, and also we have defined $b := y^2 dx^+\wedge d\psi$ (see \aammTWOFORMBBBBB), see \MaldacenaDLC\ for a discussion on field theory and the dynamics of a particle with fixed $p_-$ on the metric \aallagainspecial, where $p_-$ is the momentum that is canonically conjugate to $x^-$. The T-dual of the spacetime \aallagaindiscussion\ is the G\"odel (type) universe given in \aatt, or equivalently, in \forminvariantuone\ which we write again below
\eqn\godeluniversetdual{ds^2 = -\left[dt - r^2\left(\mu_- \omega_3 + \mu_+\bar\omega_3\right)\right]^2 + dr^2 + r^2d\Omega_3^2 + dq^2,
}
where $\mu_\pm = (\mu \pm 1)/4$, and $\omega_i$ are the left invariant one-forms and $\bar\omega_i$ are the right invariant one-forms of the unit three sphere with metric $d\Omega_3^2$.

We showed that depending on the sign of the deformation coupling single-trace $T{\bar T}$ deformation either speeds up or slows down the intrinsic rotation (or vorticity) of the universe. That is, the holographic screen \bbaa\ is either moved radially into the bulk or out to infinity relative to its original position (which is $r_{\cal H} = l_s$ in string units). This naturally suggests or shows that in holography (single-trace) $T\bar T$ deformation can be viewed, in general, as either moving the holographic boundary into the bulk or washing it away to infinity. This is already obvious and clear from Fig. 1. In AdS this breaks the spacetime conformal symmetry. We noted that moving timelike boundary into bulk creates a curvature singularity. Thus, the timelike boundary ceases to exist. The creation of the singularity is understood or noted in the deformed spacetime theory by the energies of some states becoming imaginary.

It is interesting to note that, in general, cosmological models or spacetimes, which are flat or nearly so, do not have (observer independent) timelike boundaries, and hence, as in the G\"odel (type) universe, single-trace $T{\bar T}$ deformation is not expected to generate a curvature singularity (in particular upon moving the boundaries into the bulk). Therefore, it is not expected to involve inconsistencies and pathological results. It may be useful in constructing (physical) cosmological universes and to gain insights into black hole singularity.

In a future work, we hope to study the string theory spectrum in the plane wave background \aall. In particular, we would like to study the properties of the theory as a function of the parameter $\mu$ and better understand the constraint \zzaa. We also would like to study (single trace) $T{\bar T}$ deformation  in the presence of fermions \refs{\Baggio, \ \Chang}. In particular we would like to know whether it can be understood in terms of fermionic T-dualities and shift \refs{\BerkovitsB,\ \Osten}.  We also hope to study the causal structure of the single-trace $T{\bar T}$ deformed G\"odel (type) universe with black hole \HerdeiroMM. This can teach us also the causal structure of the related Kerr spacetime under (single-trace) $T{\bar T}$ deformation. We plan also to study the asymptotic symmetry algebra of the background \summarymetric\ with a black hole.

\bigskip\bigskip

\noindent{\bf Acknowledgements:} I thank Nava Gaddam for reading the draft and useful comments. I also thank Nadav Drukker and David Kutasov for useful comments and email correspondence. I thank also the String and Quantum Gravity group at ICTS for their valuable comments and suggestions. This work is supported by the Department of Atomic Energy under project no. RTI4001.

\appendix{A}{Bosonic T-duality}

We briefly review T-duality in Type II bosonic string theory. See \refs{\Buscher\BuscherB-\Rocek, \ \BerkovitsB} for detailed discussions. 

Consider the string sigma model action
\eqn\zzza{S := \int d^2 z{\cal L}, \quad {\cal L} = \partial X^m \Sigma_{mn}(X) \bar\partial X^n.
}
The background fields are organized as
\eqn\zzzaa{
\quad \Sigma_{mn} := G_{mn} + B_{mn}, \quad G_{mn} = G_{nm}, \quad B_{mn} = -B_{nm},
}
where $G_{mn}$ are the components of the metric $G$ and $B_{mn}$ are the components of the Kalb-Ramond NSNS two-form $B$ field. We omitted a term linear in the dilaton $\Phi$ field. The dilaton enters at the quantum level, see, \eg, \FDavidMM.

We assume the background fields $\Sigma_{mn}$ and $\Phi$ are independent of $X^1$. Thus, the transformation
\eqn\zzzb{X^1 \to X^1 + c,
}
is a symmetry of the action, where $c$ is a constant. Therefore, we can couple the theory to a gauge field $A_a$ on the worldsheet. We replace $\partial X^1 \to \partial X^1 + A$ and $\bar\partial X^1 \to \bar\partial X^1 + {\bar A}$. The gauged Lagrangian density is
\eqn\zzzc{{\cal L} = \partial X^m \Sigma_{mn} \bar\partial X^n + \partial X^m\Sigma_{m1} {\bar A} + A \Sigma_{1n} \bar\partial X^n + A\Sigma_{11} {\bar A} + \chi  (\partial {\bar A} - \bar\partial A), 
}
where the field $\chi$ is a Lagrange multiplier. The equation of motion of $\chi$ forces $A_a$ to be the derivative of a scalar, \ie, $F_{ab} = \partial_a A_b - \partial_bA_a = 0$. Therefore, we get back the original same action after fixing the gauge and integrating out $\chi$. We can fix the gauge \eg, by assuming $X^1 = 0$. In general, however, $F_{ab} = 0$ does not necessarily imply the gauge field potential $A_c$ can be written globally as the derivative of a scalar field. That is, it does not necessarily imply the gauge potential is exact globally, \ie\ there could be obstructions. In general, a flat gauge field, \ie, $F_{ab} = 0$, only implies the gauge field is a pure gauge which is, by definition, a local statement.

We assume $X^1$ is periodic and compact, \ie, at fixed worldsheet time it parametrizes a circle. Thus, the symmetry \zzzb\ is a global $U(1)$ (zero-form) symmetry. See \refs{\Buscher\BuscherB-\Rocek, \ \Alvarez, \ \BerkovitsB} for discussions in the case in which $X^1$ is not compact, and in the case in which T-duality does not apply or is not an exact symmetry.

We fix the gauge by assuming $X^1 = {\rm const.}$. This gives
\eqn\zzzd{S = \int d^2z {\cal L}, \quad {\cal L} = \partial X^{\hat m} \Sigma_{{\hat m}{\hat n}} \bar\partial X^{\hat n} + \partial X^{\hat m}\Sigma_{{\hat m}1} {\bar A} + A \Sigma_{1{\hat n}} \bar\partial X^{\hat n} + A\Sigma_{11}{\bar A} + \chi  (\partial {\bar A} -  \bar \partial A),
}
 where $\hat m$ takes all values of $m$ except $m = 1$. Integrating out the field $\chi$, \ie, eliminating it using its equation of motion, gives back \zzza. The equation of motion of $\chi$ is $F_{ab} = 0$. This is solved by $A = \partial\chi^1$ and ${\bar A} = {\bar\partial}\chi^1$, where $\chi^1$ is identified with $X^1$. The first de Rham cohomology group of $S^1$ (or punctured plane) is $H^1(S^1) = \IR$.\foot{The $r$th de Rham cohomology group $H^r(M)$ of a differentiable manifold $M$ is defined by the quotient $Z^r(M)/B^r(M)$, where $Z^r(M)$ is the set of closed $r$ forms and $B^r(M)$ is the set of exact $r$ forms. Each set is a group.} Thus, in general a one form on $S^1$ is not exact.\foot{Therefore, the gauge field $A_b$ in general can have non-trivial holonomies around (non-contractible) loops (encircling a point). The non-triviality of a holonomy is in general quantified or measured by the curvature. Thus, in the new spacetime we may encounter, for example, a conical (or an orbifold) singularity. Therefore, in general in a situation in which we allow the gauge field $A_b$ to possess a non-trivial holonomy, the gauging procedure (below) is (viewed as) an orbifolding, see \refs{\Rocek,\ \Alvarez}.} However, since $X^1$ is compact and thus it winds around in either directions integer times, we can still write the gauge potential on $S^1$ as the gradient of a compact scalar and make the above identifications. This will get us back to the original action. Note $A,\ {\bar A}$ are not exact however they are closed and constrained to have integer periods or holonomies, \ie, $\oint A$ and  $\oint {\bar A} $ must be integer multiplies of $2\pi$. This is achieved by adjusting the period of the Lagrange multiplier $\chi$, see \refs{\Rocek,\ \Alvarez}.
 
 If we instead perform integration by parts, it leads to
\eqn\zzze{S' = \int d^2z {\cal L}', \quad {\cal L}' = \partial X^{\hat m} \Sigma_{{\hat m}{\hat n}} \bar\partial X^{\hat n} + A \Sigma_{11} {\bar A} - \left(\partial\chi - \partial X^{\hat m}\Sigma_{{\hat m}1} \right){\bar A} + A\left(\Sigma_{1{\hat n}} \bar\partial X^{\hat n} + \bar \partial \chi\right).
}
Integrating out the gauge fields $A$ and $\bar A$ using their equations of motion gives the T-dual Lagrangian density
\eqn\zzzf{{\cal L}' = \partial X^{\hat m} \sigma_{{\hat m}{\hat n}} \bar\partial X^{\hat n} + \partial \chi \sigma_{\chi {\hat m}} {\bar\partial  X}^{\hat m} + \partial X^{\hat m} \sigma_{{\hat m}\chi}\bar\partial\chi + \partial \chi  \sigma_{\chi\chi}  \bar \partial \chi,
}
where
\eqn\zzzg{ \sigma_{\chi\chi} = {1\over \Sigma_{11}}, \quad \sigma_{\chi\hat m} =  {\Sigma_{1\hat m}\over \Sigma_{11}}, \quad \sigma_{{\hat n}\chi} =  -{\Sigma_{{\hat n} 1}\over \Sigma_{11}}, \quad \sigma_{{\hat m}{\hat n}} = \Sigma_{{\hat m}{\hat n}} - {\Sigma_{{\hat m}1}\Sigma_{1{\hat n}}\over \Sigma_{11}}.
}
The Jacobian factor introduced in the path integral measure by the change of variables leads to a shift of the dilaton. The new dilaton $\phi$ is
\eqn\zzzh{\phi = \Phi -{1\over 2}\log \left(\Sigma_{11}\right), \quad \Sigma_{11} = G_{11}.
}
The new background fields are arranged as
\eqn\zzzi{\sigma_{{\hat m}{\hat n}} = g_{{\hat m}{\hat n}} + b_{{\hat m}{\hat n}}, \quad g_{{\hat m}{\hat n}} = g_{{\hat n}{\hat m}}, \quad b_{{\hat m}{\hat n}} = -b_{{\hat n}{\hat m}},
}
and similar expressions for $\sigma_{\chi{\hat n}},\ \sigma_{{\hat n}\chi}, \ \sigma_{\chi\chi}$. Thus, the components of the new metric $g$ and two-form $b$ field are given in terms of those in \zzzaa\ by the relations
\eqn\zzzj{g_{\chi\chi} = {1\over G_{11}}, \quad g_{\chi{\hat m}} = {B_{1{\hat m}}\over G_{11}}, \quad g_{{\hat n}\chi} = -{B_{{\hat n} 1}\over G_{11}}, \quad g_{{\hat m}{\hat n}} = G_{{\hat m}{\hat n}} -{1\over G_{11}}\left(G_{{\hat m}1}G_{1{\hat n}} + B_{{\hat m}1}B_{1{\hat n}}   \right),
}
and
\eqn\zzzk{b_{\chi{\hat m}} = {G_{1{\hat m}}\over G_{11}}, \quad b_{{\hat n}\chi} = -{G_{{\hat n} 1}\over G_{11}}, \quad b_{{\hat m}{\hat n}} = B_{{\hat m}{\hat n}} -{1\over G_{11}}\left(G_{{\hat m}1}B_{1{\hat n}} + B_{{\hat m}1}G_{1{\hat n}}   \right).
}
The relations \zzzh, \zzzj\ and \zzzk\ are also known as the Buscher rules.

Note that in general we are free to multiply the Lagrange multiplier $\chi$ in \zzzf\ by a field independent factor, \eg, to adjust its period.

We note that in the case in which $\Sigma_{11}$ has zero scaling dimensions the operator $A{\bar A}$ is exactly marginal. It does not acquire anomalous dimensions since the fields $A$ and $\bar A$ are not dynamical. Thus, we can deform \zzze\ by adding $A{\bar A}$. This gives
\eqn\zzzl{S'' = \int d^2z {\cal L}'', \quad {\cal L}'' = \partial X^{\hat m} \Sigma_{{\hat m}{\hat n}} \bar\partial X^{\hat n} + A \Sigma'_{11} {\bar A} - \left(\partial\chi - \partial X^{\hat m}\Sigma_{{\hat m}1} \right){\bar A} + A\left(\Sigma_{1{\hat n}} \bar\partial X^{\hat n} + \bar \partial \chi\right),
}
where 
\eqn\zzzjx{\Sigma'_{11} = \Sigma_{11} + \lambda,
} 
and $\lambda$ is the deformation coupling.

\appendix{B}{The single trace $T{\bar T}$ operator}

We briefly review string theory on $AdS_3$. We also give the vertex operator that defines the single trace $T{\bar T}$ deformation. We primarily follow the discussion presented in \KutasovME.

We consider bosonic string theory on $AdS_3$. It is described by the Wess-Zumino-Witten (WZW) sigma model on the group manifold $SL(2, R)$. The worldsheet action possesses the affine current (Kac-Moody) Lie alegrba $sl(2)\times sl(2)$. The chiral currents obey the OPE
\eqn\qqqzz{J^a(z_1)J^b(z_2) \sim {k\over 2}{\eta^{ab}\over (z_1 - z_2)^2} + {f^{ab}_c\over z_1 - z_2}J^c(z_2),
}
where $\eta^{ab}$ is the Killing metric, $f^{ab}_c$ are the structure constants and $k$ is the level of $sl(2)$. We will assume the level $k$ is sufficiently large. Similar expressions for the ${\bar J}^a$. The worldsheet stress-energy-momentum tensor is given by the Sugawara construction
\eqn\qqqzza{T^{\rm ws}(z) = {1\over k - 2 }\eta_{ab} J^aJ^b(z).
}
Similar expression for the ${\bar T}(\bar z)$. 

Physical operators are represented by vertex operators. The vertex operators and in general primary operators are eigenfunctions of the Laplacian on AdS. The eigenvalues give the worldsheet scaling dimensions. For a vertex operator to create a physical state it must have weight $(1, 1)$\foot{In general after multiplying it by an appropriate vertex operator from the internal CFT.}, and thus, it must be marginal. Most important vertex operators are in general exactly marginal.

The function 
\eqn\qqqzzb{\Phi_h(z; x) = {1\over \pi}\left(|\gamma - x|^2e^{\phi} + e^{-\phi}\right)^{-2h},
}
is an eigenfunction of the Laplacian operator in the coordinates $( \phi, \gamma, {\bar\gamma})$. The variables $(x, {\bar x})$ are auxiliary and labels points on a two dimensional manifold. Thus, we regard them as position coordinates on the auxiliary space.

The primary operator $\Phi_h(z; x)$ has the worldsheet scaling dimensions $(\Delta_h, {\bar\Delta}_h)$
\eqn\zzzscale{\Delta_h = \bar\Delta_h = {-h(h - 1)\over k - 2 }.
}
In momentum space, see \Asrat, we have 
\eqn\qqqzzzx{\Phi_h(z; p) = {2^{-2h + 2}\over \Gamma(2h)}\cdot e^{i{\vec p}\cdot {\vec\gamma}}\cdot e^{2(h - 1)\phi}\cdot \alpha^{2h - 1}\cdot K_{-2h + 1}(\alpha), \quad \alpha := |{\vec p}|e^{-\phi},
}
where $K_\nu$ is the modified Bessel function of the second kind and ${\vec p}\cdot {\vec\gamma} = {\bar p}\cdot{ \gamma} + {p}\cdot{ \bar \gamma}$. $K_\nu$ has the property that $K_{-\nu} = K_\nu$ for all $\nu$. Near $\alpha = 0$, we have
\eqn\qqmacd{\eqalign{K_{i\nu}(\alpha) & \approx -\log(\alpha/2)\left(1 - {1\over 6}\nu^2\log^2\left(\alpha/2\right)\right) + {\cal O}(\nu^4), \quad \nu \geq 0\cr
K_\nu(\alpha) & \approx {1\over 2}\Gamma(\nu)\left({\alpha\over 2}\right)^{-\nu}, \quad {\rm Re (\nu)} > 0.
}
}
We note using the property of $K_{\nu}$ that 
\eqn\qqssrrttuuvv{G_{-h + 1}(z; p) = {\Gamma(2h)\over \Gamma(-2h + 2)} G_h(z; p),
}
where $G_\nu$ is defined as
\eqn\qqssrrttuuvvRELATIONNN{
G_{\nu}(z;p) := \left({|{\vec p}|^2\over 4}\right)^{-\nu}\Phi_{\nu}(z; p).
}
Thus, for real $h$ we only need to know $\Phi_h(z; x)$ for $h \geq 1/2$. For complex $h = {1/2} + is$, it relates $s > 0$ with $s < 0$. Thus, the operators $\Phi_{1/2 + is}$ and $\Phi_{1/2 - is}$ are not independent. They are related as
\eqn\qqssrrttuuvvRELATION{G_{{1\over 2} - is}(z; p) = {\Gamma(1 + 2is)\over \Gamma(1 - 2is)}G_{{1\over 2} + is}(z; p).
}

In the large $\phi$ limit we find
\eqn\zzzzxxx{\Phi_h(z; p) = {1\over 2h - 1}\cdot e^{i{\vec p}\cdot {\vec\gamma}}\cdot e^{2(h - 1)\phi} = {1\over 2j + 1}e^{2j\phi + i{\vec p}\cdot {\vec\gamma}}, \quad {\rm Re}(h)  > {1\over 2}.
}
where $j = h - 1$. Using the series representation of $K_\nu(\alpha)$, in the small $\alpha$ limit, the first few relevant or dominant terms are
\eqn\zzzzxxxa{\Phi_h(z; p) = {1\over 2h - 1}\cdot e^{i{\vec p}\cdot {\vec\gamma}}\cdot e^{2(h - 1)\phi} + {\Gamma(-2h + 1)\over \Gamma(2h)} \cdot 4^{-2h + 1}\cdot |{\vec p}|^{4h - 2}\cdot e^{i{\vec p}\cdot {\vec\gamma}}\cdot e^{-2h\phi} + \cdots.
}
We rewrite this as
\eqn\zzzzxxxaREW{\Phi_h(z; p) = {1\over 2h - 1}\left(e^{2(h - 1)\phi} + R_h \cdot e^{-2h\phi} \right) e^{i{\vec p}\cdot {\vec\gamma}} + \cdots, \quad R_h =  {\Gamma(-2h + 1)\over \Gamma(2h - 1)}\left( |{\vec p}|^2\over 4\right)^{2h - 1}.
}

Note that in momentum space two point function of operators ${\cal O}_{h, h} := {\cal O}$ of conformal dimensions $(h, h)$ is $\pi\cdot R_h/(2h - 1)$, see \Asrat. In position space the operators are normalized such that $\langle{\cal O}(x){\cal O}(y) \rangle = 1/|x - y|^{4h}$. Thus, in the large $\phi$ limit
\eqn\thedict{
\Phi_h(z; p) = {1\over \pi}\left({2\pi\over 2\Delta - d}\cdot \rho^{-(d - \Delta)} + \langle{\cal O}(p){\cal O}(-p)\rangle \cdot \rho^{-\Delta}\right) e^{i{\vec p}\cdot {\vec\gamma}}, 
}
here $d = 2$, $\Delta = 2h$ and $\rho = e^{\phi}$. We hope to investigate the analog of \thedict\ in the theory \summaryaction\ in a future work.

We note that in the case $h = 1/2$ the two terms are equivalent and $R_h = -1$. For exact result one can use \qqqzzzx\ and the first equation in \qqmacd. We get (see also \HenneauxMMT\ for a related discussion)
\eqn\atonehalf{\eqalign{
\Phi_{1/2}(z; p) & = -\log\left(|{\vec p}|e^{-\phi}/2\right)\cdot e^{-\phi}\cdot e^{i{\vec p}\cdot {\vec \gamma}},\cr
& = {1\over \pi}\left[\pi \log\left(2 \cdot \rho \right) - {\pi\over 2} \log\left(|{\vec p}|^2/4\right)\right] \cdot \rho^{-1} \cdot e^{i{\vec p}\cdot {\vec \gamma}},
}
}
here $\rho = e^{\phi}$. In the large $\phi$ limit $\Phi_{1/2}$ goes to zero. In general in the case $2h - 1 = n \in \Z_+$, $R_h$ is divergent since the gamma function has simple poles at negative integers and it needs to be regularized, see \Asrat\ for further discussion.

We also note that for ${\rm Re}(h) > 1$ $\Phi_h$ is divergent. The natural norm $\int |\Phi_h|^2 \Omega_3 = {1\over 2}\int d\phi d\gamma d\bar\gamma e^{2\phi} |\Phi_h|^2$ implies $\Phi_h$ is non-normalizable for ${\rm Re}(h) > 1/2$ and $h = 1/2$, where $\Omega_3$ is the invariant volume element. Therefore, for ${\rm Re}(h) > 1/2$ and $h = 1/2$ it corresponds on the boundary to a local operator. For $h = 1/2 + is$ with $s > 0$, the norm ${1\over 2}\int d\phi d\gamma d\bar\gamma e^{2\phi} \Phi^*_h \cdot \Phi_{h'} \propto \delta^{(2)}(x - x')\cdot \delta(s - s')/s^2$, see \zzzzxxxa.\foot{Note that one can normalize the operator by some function of only $h$ such that the norm is $\delta^{(2)}(x - x')\cdot \delta(s - s')$.}  Thus, in this case $\Phi_h$ is delta-function normalizable. The operators $\Phi_h$ with $h = {1\over 2} + is$ belong to the principal continuous representations of $SL(2, R)$. They correspond to (normalizable) states in the boundary CFT.

The global chiral symmetry is generated by the zero modes. On functions they are represented by the differential operators  
\eqn\qqzzc{J^-_0 = \partial_\gamma, \quad J^3_0 = \gamma\partial_\gamma - {1\over 2}\partial_\phi, \quad J^+_0 = \gamma^2\partial_\gamma - \gamma\partial_\phi - e^{-2\phi}\partial_{\bar\gamma}.
}
Similar expressions for the ${\bar J}^a_0$. Near the (conformal) boundary $\phi = +\infty$, we note, \eg\ see \zzzzxxx, that
\eqn\qqzzc{J^-_0 \equiv {\cal D}^-_0 := -\partial_x, \quad J^3_0 \equiv {\cal D}^3_0 := -x\partial_x - j, \quad J^+_0 \equiv {\cal D}^+_0 := -x^2\partial_x - 2 x j,
}
where $j = h - 1$. They generate global conformal transformations on the auxiliary space $(x, {\bar x})$. Thus, we interpret the auxiliary space as the boundary spacetime and $(x, {\bar x})$ as the boundary spacetime coordinates.

The primary operator $\Phi_h$ has spacetime scaling dimensions $(h, h)$. In particular the primary operator $\Phi_1(z; x)$ has spacetime scaling dimensions $(1, 1)$. Near the boundary of AdS
\eqn\qqzzsstt{\Phi_1(z; x) = \delta^{(2)}(\gamma - x),
}
see \eg\ \qqqzzzx\ or \zzzzxxx. Thus, we write near the boundary of AdS
\eqn\vertexADS{\Phi_{j + 1}(z; x) = {1\over 2j + 1}\delta^{(2)}(\gamma - x)e^{2j\phi}, \quad {\rm Re}(j) > -{1\over 2}.
}
Note, this implies, near the boundary the dominant contribution comes from $\gamma = x$. Thus, we have from \qqqzzb\ $\Phi_{h} \approx e^{2h\phi}$.

For convenience we now organize the chiral currents. We first introduce a near-boundary worldsheet current by taking some linear combination. Consider the chiral current
\eqn\qaqaqaxx{J_0(z; y): = -J^+_0(z; -y) = -e^{y{\cal D}_0^-}J^+_0(z; 0) e^{-y{\cal D}_0^-} = -J^+_0(z) + 2y J^3_0(z) - y^2J^-_0(z),
}
where $J^+_0(z) := J^+_0(z; 0)$. Now define the current
\eqn\qaqaqaxTx{J(z; y): = -J^+(z) + 2y J^3(z) - y^2J^-(z).
}
We note that $J$ has spacetime scaling dimension $(-1, 0)$, see \qqzzc. Similarly we then note that 
\eqn\qqzzcv{ J^3(z; -x) = {1\over 2}\partial_xJ(z; x), \quad J^-(z; -x) = -{1\over 2}\partial^2_xJ(z; x) = J^-(z).
}
Similar expressions for the ${\bar J}^a({\bar z}; -{\bar x})$. 

The worldesheet stress-tensor \qqqzza\ is given in terms of the worldsheet current $J$ by
\eqn\qqqzzzxxx{T^{\rm ws}(z) = {1\over 2(k - 2)}\left[J\partial^2_xJ - {1\over 2}(\partial_x J)^2\right].
}
Here we set $x = 0$. Similar expression for the ${\bar T}^{\rm ws}({\bar z})$.

Primary fields (near the boundary) are defined with respect to $J$ via the OPE
\eqn\qqzzcvv{J(z_1; x_1)\Phi_h(z_2; x_2) \sim {2h(x_2 - x_1) + (x_2 - x_1)^2\partial_{x_2}\over z_1 - z_2}\Phi_h(z_2; x_2), 
}
Similar OPE expression with ${\bar J}({\bar z}_1, {\bar x}_1)$.

The near boundary gravition or spin 2 vertex operator polarized along the boundary of AdS and corresponding to diffeomorphisms that preserve the asymptotic form of AdS is given by
\eqn\qqqzzzxxu{T(z; x) = {1\over 2(k - 2)}\left(\partial_x J \partial_x\Phi_1 + 2\partial^2_x J\Phi_1\right){\bar J}.
}
In three spacetime dimensions the graviton has no propagating degrees of freedom. Thus, the graviton resides solely on the boundary. The spacetime energy-momentum tensor is obtained by integrating the (near-boundary) graviton vertex operator
\eqn\qqqzzzxxuU{T(x) = \int d^2 z T(z; x).
}
Similar expression for ${\bar T}({\bar x})$.

The dilaton or spin zero vertex operator (near the boundary) is given by 
\eqn\qqzzcvxx{D(z; x) = \left(\partial_x J\partial_x + 2\partial^2_x J\right)\left(\partial_{\bar x} {\bar J}\partial_{\bar x} + 2\partial^2_{\bar x} {\bar J}\right)\Phi_1.
}
The boundary dilaton or spin zero operator is then
\eqn\qqzzcvxxT{D(x) = \int d^2z D(z; x).
}
It obeys the relation 
\eqn\qqqzzzxx{\int d^2 x D(x, {\bar x}) = 4\int d^2 z J^-{\bar J}^-(z, {\bar z}).
}
This follows by changing order of integrations and performing integration by parts. In the Wakimoto realization of the $sl(2)$ Lie algebra, the conserved current $J^- := \beta$, see Appendix C. $D(x, {\bar x})$ has the spacetime scaling dimensions $(2, 2)$. The equality \qqqzzzxx\ defines the single trace $T{\bar T}$ deformation. See Appendix C for more discussion on this.

\appendix{C}{Single trace $T{\bar T}$ deformation}
We briefly review the single trace deformation and give the deformed spacetime spectrum in a given winding or twisted sector. We mainly follow the discussion presented in \Asrat.

Bosonic string theory on $AdS_3$ can be equivalently described in the Wakimoto realization of $SL(2, R)$ by the worldsheet action 
\eqn\appcaa{S = {1\over 2\pi} \int d^2z \left(\partial\phi{\bar\partial}\phi  - {2\over \alpha_+}\phi R^{(2)} + \beta{\bar\partial}\gamma + {\bar\beta}\partial{\bar\gamma} - \beta\bar\beta e^{-{2\over \alpha_+}\phi}\right),
}
where $\alpha_+ = \sqrt{2(k - 2)}$, $k$ is the level and $R^{(2)}$ is the curvature of the wordsheet. At $\phi = +\infty$, the null coordinates $\gamma = x - t$ and $\bar\gamma = x + t$. Therefore, the zero mode of the current $J^{-}$ associated with the Killing vector field $\partial_\gamma$ is related to the spacetime energy $E$ and momentum $P$. The fields $\beta$ and $\bar\beta$ can be identified with the background fields $A$ and $\bar A$ in \zzzl. Since the operator $e^{-{2\over \alpha_+}\phi}$ has zero worldsheet scaling dimensions, as we noted in \zzzl, we can deform the action \appcaa\ by adding $\beta\bar\beta$. This gives
\eqn\appcac{S_\lambda = {1\over 2\pi} \int d^2z \left[\partial\phi{\bar\partial}\phi   + \beta{\bar\partial}\gamma + {\bar\beta}\partial{\bar\gamma} - \beta\bar\beta\left(e^{-{2\over \alpha_+}\phi} - \lambda\right)\right].
}
We omitted the dilaton term. The coupling $\lambda$ is dimensionless. Integrating out the fields $\beta$ and $\bar\beta$ using their equations of motion we get
 \eqn\appcad{S_\lambda = {1\over 2\pi} \int d^2z \left(\partial\phi{\bar\partial}\phi  + {e^{{2\over \alpha_+}\phi}\over 1 - \lambda e^{{2\over \alpha_+}\phi}}\partial{\bar\gamma}{\bar\partial}\gamma \right),
}
which is in agreement with \aagg. The dilaton is determined by requiring worldsheet conformal invariance (to all orders in $\alpha'$) \refs{\Hassan,\ \Araujo}.

The Hilbert space of string theory on $AdS_3$ is given by a tensor product of a short and long string sectors.  Near the asymptotic boundary, \ie, in perturbation theory in the string coupling $g_s^2 \propto 1/p \propto 1/N$, the long string sector is believed to be described in the dual boundary theory by a symmetric product $S^N({\cal M}) := {\cal M}^N/S_N$, where $p$ is the (maximal) number of long strings, $N$ is the degree of the symmetric group $S_N$ and ${\cal M}$ is the theory dual to a single long string.

In the long string sector in the case the theory is quantized on a circle the deformed spacetime spectrum is given by \Asrat,
\eqn\sypectrQQ{ {\bar E}(\lambda) = {\bar P} - {1\over 2A}\left(B + \sqrt{B^2 - 4AC}\right),
}
where
\eqn\sypectQQQ{A = {1\over 2}\lambda, \quad B = -w + \lambda {\bar P}, \quad C = w({\bar E}_0 - {\bar P}),
}
and
\eqn\sypectQQQW{{\bar E} = {R{\cal E}\over 2\pi}, \quad {\bar P} = {RP\over 2\pi}, \quad {\bar E}_0 = {\bar E}(\lambda = 0),
}
where $w$ is the winding number of the long string which counts the number of times the string winds around the compact spatial direction. We rewrite the spectrum \sypectrQQ\ as
\eqn\sypectQQQWQ{{\bar E}(\alpha) = {1\over \alpha}\left(\sqrt{1 + 2\alpha {\bar E}_0 + \alpha^2 {\bar P}^2} - 1\right), \quad \alpha = -{\lambda\over w}.
}
Note ${\bar E_0}$ is the energy of the undeformed state in the winding or twisted sector $w$ and ${\bar P}$ is its momentum. In the case $w = 1$ or untwisted sector the spectrum \sypectrQQ\ is that of a $T{\bar T}$ deformed CFT. This is in agreement with the identification that in the long string sector the single trace deformation $D(x,{\bar x})$ is equivalent to a symmetric product of the $T{\bar T}$ deformed CFT ${\cal M}$.

We note that in the $w$ twisted sector the states with ${\bar E}_0 = {\bar P}$ do not flow. That is, ${\bar E}(\alpha) = {\bar E}_0$.

\appendix{D}{The Maurer-Cartan form on the three sphere}

The three sphere $S^3$ is the group manifold of $SU(2)$. The generators of the Lie algebra $su(2)$ are
\eqn\vvvsigma{a_1 = {1\over 2}i\sigma_1, \quad a_2 = {1\over 2}i\sigma_2, \quad a_3 = {1\over 2}i\sigma_3, \quad [a_i, a_j] = -\epsilon_{ijk}a_k, \quad \epsilon_{123} = +1.
}
where $\epsilon_{ijk}$ is the totally antisymmetric tensor and $\sigma_1,\ \sigma_2$ and $\sigma_3$ are the Pauli spin matrices. They are given by
\eqn\paulisigma{\sigma_1 = \pmatrix{
0 & 1\cr
1 & 0
}, \quad \sigma_2 = \pmatrix{
0 & -i\cr
i & 0
}, \quad \sigma_3 = \pmatrix{
1 & 0\cr
0 & -1
}, \quad \sigma_i\sigma_j = \delta_{ij}I_{2\times 2} + i\epsilon_{ijk}\sigma_k, \quad \epsilon_{123} = +1.
}
Here and below a sum over repeated indices is assumed.

We parametrize the group element $g_{S^3} \in SU(2)$ as
\eqn\sutwog{g_{S^3} = e^{i\varphi\sigma_3/2}e^{i\theta\sigma_1/2}e^{i\psi\sigma_3/2} = \pmatrix{
e^{i(\varphi + \psi)/2}\cos(\theta/2) & ie^{i(\varphi - \psi)/2}\sin(\theta/2)\cr
ie^{-i(\varphi - \psi)/2}\sin(\theta/2) & e^{-i(\varphi + \psi)/2}\cos(\theta/2)
},
}
where the parameters $\varphi,\ \theta,\ \psi$ are the Euler angles. They satisfy the conditions
\eqn\sphereparam{0\leq \varphi < 2\pi, \quad 0\leq \theta < \pi, \quad -2\pi \leq \psi < 2\pi.
} 
The inverse is given by
\eqn\sutwogin{g_{S^3}^{-1} = \pmatrix{
e^{-i(\varphi + \psi)/2}\cos(\theta/2) & -ie^{i(\varphi - \psi)/2}\sin(\theta/2)\cr
-ie^{-i(\varphi - \psi)/2}\sin(\theta/2) & e^{i(\varphi + \psi)/2}\cos(\theta/2)
}.
}

The left invariant Maurer-Cartan one forms $\omega_i$ are given by
\eqn\leftone{\omega = g_{S^3}^{-1}dg_{S_3} = a_i\omega_i,
}
where
\eqn\vvvonef{\omega_1 = \cos\psi d\theta - \sin\psi\sin\theta d\varphi, \quad \omega_2 = -\sin\theta\cos\psi d\varphi - \sin\psi d\theta, \quad \omega_3 = d\psi + \cos\theta d\varphi.
}

The right invariant Maurer-Cartan one forms $\bar\omega_i$ are given by
\eqn\leftone{\bar\omega = dg_{S^3}g_{S_3}^{-1} = a_i\bar\omega_i,
}
where
\eqn\vvvonef{\bar\omega_1 = \cos\varphi d\theta + \sin\varphi\sin\theta d\psi, \quad \bar\omega_2 = \sin\theta\cos\varphi d\psi - \sin\varphi d\theta, \quad \bar\omega_3 = d\varphi + \cos\theta d\psi.
}

The metric $d\Omega_3^2$ on the unit three sphere $S^3$ is given by
\eqn\metricsphereethree{
\eqalign{d\Omega_3^2  & = -\tr\left(g_{S^3}^{-1}dg_{S_3}g_{S^3}^{-1}dg_{S_3}\right), \cr
& =  {1\over 4}\sum_{i = 1}^3\omega_i^2 = {1\over 4}\sum_{i = 1}^3{\bar\omega}_i^2,  \cr
   & = {1\over 4}\left(d\theta^2 + d\varphi^2 + d\psi^2 + 2d\varphi d\psi \cos\theta\right). 
}
}
We rewrite this as
\eqn\metricsphere{d\Omega_3^2 = {1\over 4}\left(d\theta^2 + d\varphi^2 + d\psi^2 + 2d\varphi d\psi \cos\theta\right) = d\tilde\theta^2 + \cos^2\tilde\theta d\tilde\theta_+^2 + \sin^2\tilde\theta d\tilde\theta^2_-,
}
where 
\eqn\coordpluminus{\tilde\theta = {1\over 2}\theta, \quad \tilde\theta_+ = {1\over 2}\left(\varphi + \psi\right), \quad \tilde\theta_- = {1\over 2}\left(\varphi - \psi\right).
}

We also note that
\eqn\someidenti{{1\over 4}\mu(\bar\omega_3 + \omega_3) + {1\over 4}(\bar\omega_3 - \omega_3) = \mu\cos^2\tilde\theta d\tilde\theta_+ + \sin^2\tilde\theta d\tilde\theta_-.
}

\appendix{E}{The fundamental domain of the coupling}
Consider the transformation 
\eqn\transinv{
\gamma \to \eta = {1\over \gamma}.
}
The dilaton becomes
\eqn\tradila{\eqalign{{e^{2\phi_\gamma}\over g_s^2} & = {1\over 1 + \gamma^2 - 2\gamma \cosh 2\tilde\theta}, \cr
& = {\eta^2\over 1 + \eta^2 - 2\eta \cosh 2\tilde\theta} = \eta^2\left(e^{2\phi_\eta}\over g_s^2\right).
}}
The $B$ field becomes
\eqn\trabb{\eqalign{B_{01}^\gamma & = -{1\over 2}k\alpha' {e^{2\phi_\gamma}\over g_s^2}(\gamma - \cosh 2\tilde\theta),\cr
& = -{\eta^2\over 2}k\alpha'{e^{2\phi_\eta}\over g_s^2}\left({1\over \eta} - \cosh 2\tilde\theta\right).
}
}
A physically equivalent $B$ field is
\eqn\trabbequiv{\eqalign{B_{01}^{'\gamma} & = B_{01}^{\gamma} + {1\over 2}k\alpha' \cdot {1\over \gamma}, \cr
& =  -{\eta^2\over 2}k\alpha'{e^{2\phi_\eta}\over g_s^2}\left({1\over \eta} - \cosh 2\tilde\theta\right) + {1\over 2}k\alpha' \eta,\cr
& =  -{\eta^2\over 2}k\alpha'{e^{2\phi_\eta}\over g_s^2}\left({1\over \eta} - \cosh 2\tilde\theta - {g_s^2 e^{-2\phi_\eta}\over \eta }\right),\cr
& = -{\eta^2\over 2}k\alpha' {e^{2\phi_\eta}\over g_s^2}(-\eta + \cosh 2\tilde\theta) = -\eta^2 B_{01}^\eta.\cr
}
}
We now make the following change of variables
\eqn\changetraninv{\tilde\varphi \to \tilde\chi = \eta \tilde\varphi , \quad \tilde\psi \to \tilde\xi = -\eta\tilde\psi.
}
The angle variable $\tilde\psi(\tau, \sigma)$ has the period
\eqn\tranperiodinv{
\tilde\psi(\tau, \sigma + 2\pi) = \tilde\psi(\tau, \sigma) + 2\pi(1 - \gamma).
}
Therefore,
\eqn\tranperiodinv{\eqalign{
\tilde\xi(\tau, \sigma + 2\pi) & = -\eta\tilde\psi(\tau, \sigma + 2\pi),\cr
& =  -\eta\tilde\psi(\tau, \sigma) - \eta \cdot 2\pi \left(1 - {1\over \eta}\right),\cr
& = \tilde\xi(\tau, \sigma) + 2\pi (1 - \eta).
}
}
Thus, the transformations and/or redefinitions \transinv\ and \changetraninv\ leave the metric and $B$ field invariant. Rescaling the string coupling as
\eqn\strrescal{g_s^2 \to \eta^2 g_s^2,
}
further ensures the dilaton remains the same. Therefore, we can restrict $\gamma$ to take values in the fundamental domain $|\gamma| \leq 1$. 

\listrefs

\bye